\documentclass{article}
\pdfpagewidth  = \paperwidth
\pdfpageheight = \paperheight

\usepackage{graphicx}
\usepackage{epstopdf}
\usepackage{url}
\usepackage{booktabs}
\usepackage{algorithm}
\usepackage{algorithmic}
\usepackage{listings}
\usepackage{mathtools}
\usepackage{filecontents}

\begin{document}
\title{The Anatomy of a Search and Mining System for\\Digital Archives}
\date{}
\maketitle
\begin{center}
\author{
  Martyn Harris, Mark Levene, Dell Zhang\\
  
      Department of Computer Science and Information Systems\\
      Birkbeck, University of London\\
      London WC1E 7HX, UK\\
      martyn, mark, dell @dcs.bbk.ac.uk
      \linebreak 
      \linebreak	
  \and
  Dan Levene\\
  
      History, School of Humanities\\
      University of Southampton\\
      Southampton SO17 1BF, UK\\
     d.levene@soton.ac.uk\\
   
}
\end{center}
%\pagebreak
%\tableofcontents
%\pagebreak
%\listoffigures
%\listoftables

\begin{abstract}
Samtla (Search And Mining Tools with Linguistic Analysis) is a digital humanities system designed in collaboration with historians and linguists to assist them with their research work in quantifying the content of any textual corpora through approximate phrase search and document comparison. 
The retrieval engine uses a character-based $n$-gram language model rather than the conventional word-based one so as to achieve great flexibility in language agnostic query processing. \\
The index is implemented as a space-optimised character-based suffix tree with an accompanying database of document content and metadata. 
A number of text mining tools are integrated into the system to allow researchers to discover textual patterns, perform comparative analysis, and find out what is currently popular in the research community.\\
Herein we describe the system architecture, user interface, models and algorithms, and data storage of the Samtla system. We also present several case studies of its usage in practice together with an evaluation of the systems' ranking performance through crowdsourcing.
\end{abstract}
\pagebreak

\section{Introduction}
Digital textual representations of original historic documents are being made available thanks to the work of digital archiving projects \cite{Rommel2007}. As the wealth of digitised textual data increases it becomes more important to provide the appropriate tools for analysing and interrogating the sources. Humanities researchers are discovering how digital tools can become part of their methodology \-- even collating their source materials into a simple repository can help to speed up the access to resources which were otherwise stored in physical libraries.  However, there are still barriers to adoption, including usability and the scope of the provided tools \cite{Gibbs2012, Sweetman2012}. In addition, there is a tendency to develop systems which are language specific and rely on part of speech taggers to identify all instances of a word, regardless of its morphology, in order to provide accurate recall. Furthermore, such systems may be tied to a particular document collection, for example, the works of William Shakespeare. As a result, when developing any tool for the humanities important aspects to consider are how to index, search, compare, and apply data mining tools to domain-specific corpora represented by a collection of documents grouped together for a specific research agenda.   \\
\\
Samtla has been designed to provide a research environment that is agnostic to the document collection and can therefore be used by a wide range of research groups whose work involves analysing digital representations of original source texts. It currently supports search, browsing, and analysis of texts through approximate phrase searches, related query and document recommendations, a document comparison tool, and community features in the form of popular queries and documents. Samtla's interface adopts the flat design principle, which reflects current trends \cite{flatdesign1, flatdesign2} in user interface design in terms of user interaction and the layout of components such as tool bars and informational side panels to promote familiarity with respect to applications the user may use regularly (including browsers, music and video players, cloud storage), and legibility in terms of centralising the content by presenting it clearly to users.\\
\begin{figure}[!tbh]
	\centering
  	    \fbox{\includegraphics[width=\columnwidth]{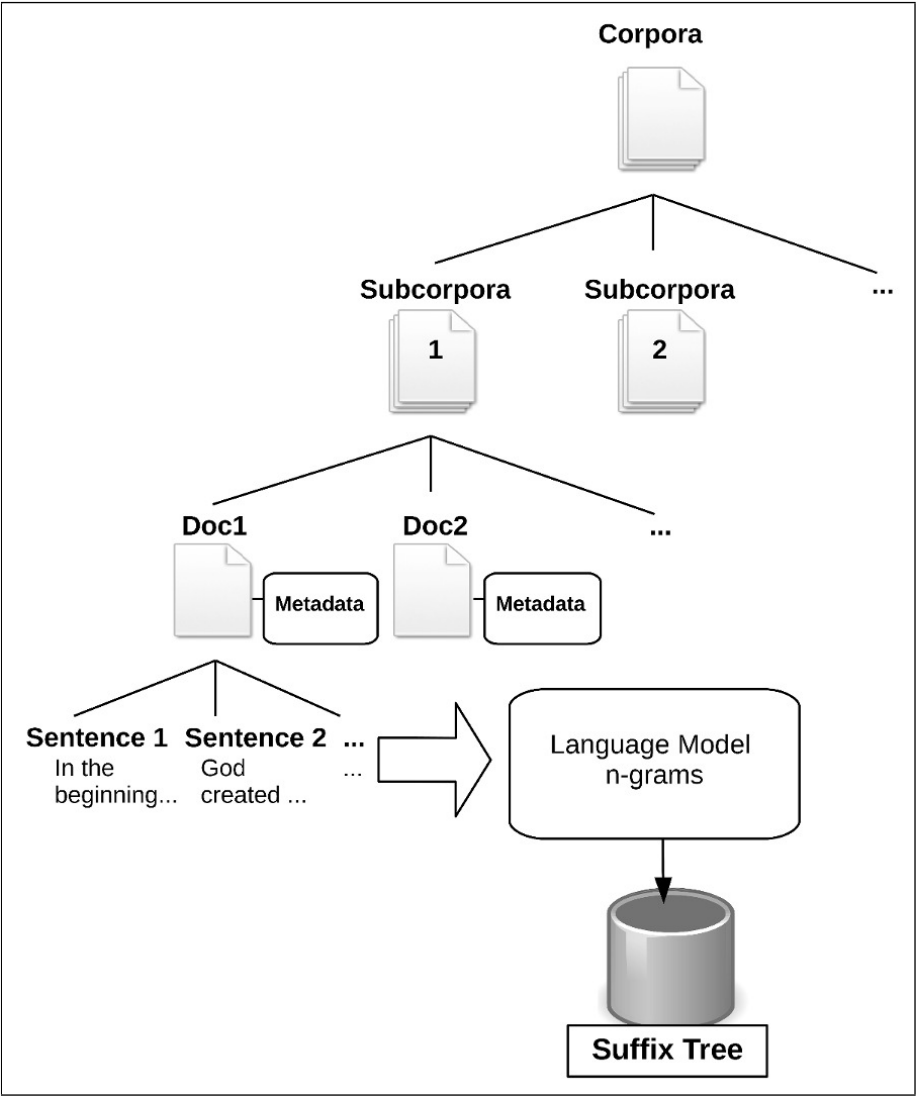}}
	\caption{The corpus typology.}
	\label{fig:typology}
\end{figure}
\\
Samtla was developed in collaboration with historians and linguists to cater for their research needs in quantifying the content of textual corpora. We have adopted a Statistical Language Model (SLM) for information retrieval \cite{Zhai2009}, and incorporated text mining tools into the system to allow researchers to go beyond a pure search and browse paradigm. Such an extended ``search and research'' model supports the discovery of patterns (known by historians as ``formulae'' or ``parallel passages'') that have significance in terms of their research goals. The ``formulae'' are reflected by textual fragments represented, for example, by set phrases or quotations. The main textual fragment is duplicated across several documents with slight variations resulting from differences in authorship, language change due to locality and time, which can manifest themselves through dialectal differences. The system has been designed to be applicable to any type of corpus in any language with little pre-processing, and to provide transparency in terms of its functionality, in order to help researchers adopt it as an integral part of their research strategy \cite{Gibbs2012, Sweetman2012}. For this purpose, our \emph{n}-gram statistical language model is character-based rather than word-based, which makes the query processing and further analysis flexible. For instance, consider the difficulty involved in indexing words from a collection of English news articles and those in the Chinese language; the latter of which has no whitespace equivalent word boundary marker as that present in English.\\
\\
Figure~\ref{fig:typology} illustrates the basic structure of a corpus and the documents contained within.  Some corpora maybe partitioned in to subcorpora, and documents may also come with metadata, either in terms of document features (title, pages, and length) or additional metadata items generated by the user group.   \\
\\
Samtla currently operates with five case study textual corpora: a collection of Aramaic texts from late antiquity, Italian and English translations of the writings of Giorgio Vasari, the Microsoft corpus of 68,000 scanned books, the King James Bible in English, and a test corpus of scanned Newspaper articles from the Financial Times as part of a pilot study in collaboration with the British Library. Samtla was developed to enable faster access to the document collection and to compliment existing methods adopted by our users for comparison and discovery of related documents and parallel text fragments.  \\
\\
In the following sections we discuss prominent tools that are currently available for researchers (Section~\ref{sec:relatedwork}), outline each of the system components described briefly above in more detail including the Samtla system architecture (Section~\ref{sec:architecture}), the statistical models underlying document scoring and recommendation, and a discussion of the chosen implementation methods used to structure and organise the data used by the system (Section~\ref{sec:datamodel}). We introduce the tools that have been implemented to help researchers browse, view documents and related metadata, and compare shared-sequences between document pairs (Section~\ref{sec:textmining}). Next, we introduce the recommendation component of the system, which leverages data collected from the user query submissions and document views to generate recommended queries and documents to help users locate interests aspects of the collection (Section~\ref{sec:community}). Section~\ref{sec:UI} provides a description of the User Interface ({\em UI}) and discusses how the various tools have been incorporated in to the interface and how we expect users to navigate through the system.  We also present case studies describing how the system is currently being used by researchers (Section~\ref{sec:casestudy}), before moving to the results of a formal evaluation using a crowdsourcing platform (Section~\ref{sec:evaluation}). Lastly, we conclude the paper with a summary of the work and future development plans (Section~\ref{sec:conclusion}).
\section{Related Work}
\label{sec:relatedwork}
There are a number of existing systems that provide state of the art tools in the Humanities. We discuss some of the systems that share common functionality, source material, or user groups with reference to Samtla.  \\
\\
The Bar-Ilan \emph{Responsa} project, established in 1963, is one of the earliest examples of Humanities researchers adopting the use of computer-based methods for search, comparison, and analysis of Hebrew texts. The corpus spans approximately three thousand years, and includes the Mishnah, Talmud, Torah, and the Bible in Aramaic \cite{Choueka1989}. The Responsa environment is packaged on a CD-ROM and provides browsing and searching the corpus using keyword and phrase search, comparison of parallel passages, author biographies, and user comment and annotation tools \cite{responsafeatures}.  The Bar-Ilan \emph{Responsa} project has made a considerable contribution to Computational Linguistics and information retrieval for Hebrew.\\
\\
The 1641 depositions project at Trinity College Library (Dublin) \cite{1641depositions} adopted IBM's \emph{LanguageWare} \cite{IBMLW} for the analysis of 31 volumes of books containing 19,010 pages of witness accounts reporting theft, vandalism, murder, and land taking during the conflicts between Catholics and Protestants in 17th century Ireland.  \emph{LanguageWare} provides text analysis tools for mining facts from large repositories of unstructured text. Its main features are dictionary and fuzzy look-up, lexical analysis, language identification, spelling correction, part-of-speech disambiguation, syntactic parsing, semantic analysis, and entity and relationship extraction.  \emph{LanguageWare} was chosen due to the complexity of the language contained in the documents, which have many spelling mistakes making analysis a complex task.  \emph{CULTivating Understanding Through Research and Adaptivity} (\emph{CULTURA}) is a project related to the 1641 depositions, launched in 2011 \cite{CULTURA}. \emph{CULTURA} provides users with tools for normalising texts containing inconsistent spelling, entity and relationship extraction present within unstructured text, and social network analysis tools for displaying the entities and relationships from metadata through an interactive user environment.\\
\\
Aside from tools developed as part of funded projects, we also see new applications for mobile and touch devices, developed specifically for exploring well known texts such as the Bible.  One such tool is \emph{Accordance} for \emph{Apple iOS}\cite{accordance}, which is a Bible study application featuring exact and flexible query search tools, a browsing, a timeline for viewing when people lived and important events that took place, and an atlas view for exploring journeys and battles.\\
\\
The systems descibed above are successful at providing the appropriate tools for specific groups of researchers to explore a particular collection of texts.  The main issue, however, is the generalisability of their systems to other text domains and natural languages.  These systems use language specific tools that operate at the word or morphological level requiring affix removal, tokenisation, lemmatisation, and part-of-speech tagging for normalising the texts and capturing all instances of a word in order to generate an accurate retrieval model. The question is whether they could be extended to languages with no clear word-level boundary markers (e.g. Chinese languages such as Mandarin), or without the complex rules necessary to identify affixes (e.g. Hebrew, Aramaic, Arabic, Italian, and Russian). Regardless of the approach adopted for text normalisation, these systems are by their nature language dependent. If we are to keep up with the volume of output generated by digitisation projects, then a new approach is necessary.  The Samtla system was designed to address the need for tools for the Digital Humantities through the creation of a flexible language-independent framework for searching, browsing, and comparing documents in a text collection that could be generalised to any document collection due to its data-driven design.  Samtla shares many of the features and tools available in the systems outlined above, however it differs in many respects due to the language-independent framework that can be extended in many novel ways without changes to the underlying system each time a new corpus is introduced. This enables a Samtla system to be deployed relatively quickly, allowing document collections and archives to be unlocked to the general public, or for research once the digitised materials are made available. The philosophy underlying the development of Samtla has been to provide the basic tools first (browse, search, and comparison), and then to develop further features through consultation with our users to discover what tools they actually need in order to be able to carry out their research.\\
\section{System Architecture}
\label{sec:architecture}
\begin{figure}[!tbh]
	\centering
		\fbox{\includegraphics[width=\columnwidth]{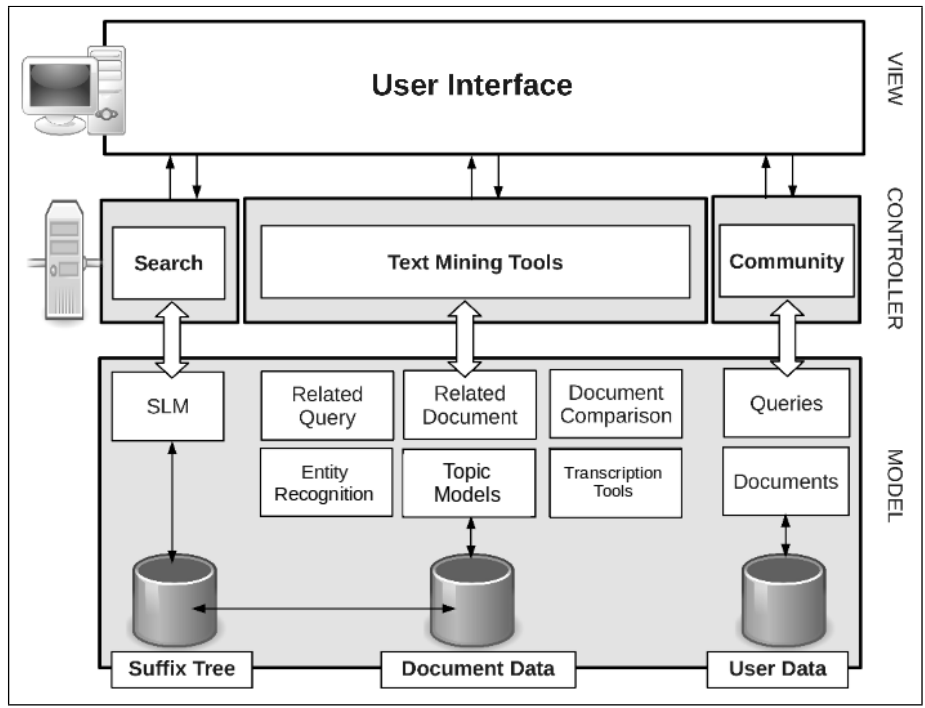}}
	\caption{The Samtla architecture.}
	\label{fig:architecture}
\end{figure}

The Samtla system is a web-based application built on a client-server architecture, providing a platform-independent solution for its deployment through a web browser.  The Samtla system operates with a single code-base, with the only corpus-dependent component being a wrapper function, which is responsible for parsing the documents or metadata to the system.  This enables the system to be data-driven and allows upgrades or changes to the functionality of Samtla to be rolled-out simultaneously to each user group.\\
\\
The client is represented by a web-based User Interface (UI), which sends requests to the server and renders the results within the browser. A central server stores all the data associated with the system, processes requests sent from the client, and responds with the appropriate data, for instance, a list of search query results.\\
\\
Samtla can be viewed as {\em a Model-View-Controller} (MVC) design pattern \cite{leff2001}, allowing the separation of the system by function.  The advantage of adopting a MVC implementation is that changes to the UI are independent of the underlying logic of the system. Therefore, due to the separation of components, introducing new features and changes to the look or functionality of the UI can be easily implemented without affecting other components.  An overview of Samtla is shown in Figure~\ref{fig:architecture}, where arrows in the diagram represent the flow of communication between the various components of the system.\\
\\
The client-side of the system is represented by the {\em view} component (i.e. the UI) providing a web-based browser interface, which allows the user to interact with the system; see Section~\ref{sec:UI} for more detail on the UI. Such interactions cause events to be triggered and picked up by the {\em controller}, representing the program logic. Technically, the client communicates with the server through the controller using URL requests, which are mapped to an appropriate function call in the {\em model}. This instigates a change to the data or the retrieval of information such as search query results or metadata.  The {\em model} returns data to the controller to process, which passes the results to the view for rendering to the UI. The controller is implemented in the Django web framework \cite{django}, which processes client HTTP requests and sends a response.  The data is passed between components using the JavaScript Object Notation (JSON) \cite{JSON} format, which allows us to store data objects that can be further processed in the browser (i.e. for dynamic rendering of HTML snippets for the search results), or static HTML fragments which are rendered directly (i.e. the raw documents). The UI is developed in Javascript with JQuery \cite{JQuery}, providing cross-browser support for the interactive elements of the interface. In addition, the system uses a number of HTML5 APIs \cite{HTML5}, including web storage for persisting the user's system preferences.\\
\\
The Samtla system libraries and data are encompassed by the {\em model} component. Samtla is written in Python and all system data is stored in SQL databases, except for the suffix tree, which is stored in {\em JSON} format and serialised to disk; see Section~\ref{sec:suffixtree} for more detail on the suffix tree component.\\
\\
The {\em model} is composed of a library of software tools that interact with the system data. The {\em search} component is responsible for answering user queries, and uses a {\em Statistical Language Model} (SLM), which is a relatively recent framework for information retrieval \cite{Zhai2009}; see Section~\ref{sec:LM} for more detail on how we make use of SLM in Samtla.  The SLM communicates with a suffix tree data structure, which is used to index the corpora that are being investigated. The suffix tree is loaded into memory at runtime for fast access. The suffix tree also provides support for the text mining tools, which are detailed in Section~\ref{sec:textmining} and include a {\em related query} feature, which recommends queries to the user based on permutations of the original query resulting from morphological or orthographic variations present in the corpus (discussed in Subsection~\ref{sec:Relatedqueries}), a {\em related document} tool, presents users with a list of similar documents to the one they are viewing (discussed Subsection~\ref{sec:Relateddocuments}), and a {\em document comparison} tool facilitates the comparison of shared-sequences between documents (discussed in Subsection~\ref{sec:doccomp}).  Lastly, the {\em community} component is responsible for logging user data, such as query submissions and document views, usage statistics reflecting the user's navigation histories through the system, and for returning to the user recommended queries and documents based on their popularity in the user community; see Section~\ref{sec:community} for more detail.\\
\\
In the following sections we discuss in detail the methods and algorithms adopted to support the current set of tools divided into search, text mining, and community support. \\
\section{Data Model}
\label{sec:datamodel}
Statistical language modelling \cite{Zhai2009} is central to Samtla's data model. It provides the foundation for Samtla's search tool allowing users to locate documents through full and partial matches to queries. Samtla's {\em Statistical Language Model} (SLM), whose details are given in Subsection~\ref{sec:LM}, is supported by a character-based {\em suffix tree} \cite{Gusfield1997}, described in detail in Subsection~\ref{sec:suffixtree}. A suffix trees is a very powerful data structure supporting fast retrieval of sequences of characters, known as {\em n-grams}, where $n$ is the length of a sequence (for our purposes, measured in characters). Samtla is unique in that it is language agnostic and can thus support a variety of languages within a single data model; we will demonstrate this in Subsection~\ref{sec:casestudy}.\\
\subsection{Statistical Language Models}
\label{sec:LM}
A SLM is a mathematical model representing the probabilistic distribution of words or sequences of characters found in the natural language represented by text corpora \cite{Rosenfeld2000, liucroft2005, Zhai2009}. Samtla is designed as a language agnostic search tool and as such uses a character-based $n$-gram SLM, rather than the more conventional word-based model. Language modelling provide Samtla with a consistent methodology for retrieving and ranking search results according to the underlying principles and structure of the language present in a corpus, which is often domain specific. Beyond that, SLMs provide a unifying model for Samtla's text mining tools described in Section~\ref{sec:textmining}.\\
\\
Statistical language modelling combined with character-level (1-gram) suffix tree nodes enable the system to be applied to multilingual corpora with very little pre-processing of the documents, unlike word-based systems. For example, languages like Hebrew, Russian, and Italian attach affixes to a root word to identify syntactic relationships. This complicates word-based retrieval models since it is necessary to capture all instances of the same word in order to produce an accurate probabilistic model. Word-based models typically require a language-dependent stemming, part-of-speech tagging, or text segmentation algorithm, however, by adopting a character-based $n$-gram model these issues can be ignored to some extent and character-based models have been shown to outperform raw word-based models, especially when the language is morphologically complex \cite{Mcnamee2004}. Furthermore, a character-based model enables the system to be applied to different language corpora, but also corpora which contain documents written in several different languages. For example, some documents in the Aramaic collection contain texts written in Hebrew, Judeo-Arabic, Syriac, Mandaic, and Aramaic, the Vasari corpus contains English and Italian documents, whereas the British Library Microsoft Corpus covers a range of languages including English, French, Spanish, Hungarian, Romanian, and Russian.\\
\\
Operationally, when the user submits a query, a list of documents is returned and ranked according to how relevant the document is to the query.  The notion of relevance refers to the users expectation of which documents should be present at the top of the ranked list, in other words, which documents the user may be looking for \cite{Zhai2009}. In Samtla we take the view that the more probable a document in the SLM sense, the more relevant it is, thus avoiding the philosophical debate on the notion of ``relevance'' \cite{Mizzaro1997}. This equates to the system retrieving the most probable documents based on the SLM representing the distribution of the $n$-grams in the corpus being searched.\\
\\
In Samtla the data comes from corpora, which consist of a collection of text documents grouped according to a specific topic, genre, demographic, or origin (e.g institution storing original versions of the digital texts).  For instance, a Bible corpus can be composed of several Bibles from different periods or translations (Wycliffe Bible, Tyndale Bible, or Thomas Young's Translation). Each Bible therefore represents an individual corpus containing a collection of documents, which represent each chapter of the given Bible. \\
\\
We generate SLMs from the corpora over the whole collection, which we call the {\em collection model}, and over each individual document, which we call the {\em document model}. A generic SLM is denoted by $M$, while the collection model is denoted by $C$ and a document model is denoted by $D$. Each SLM is generated from the $n$-grams extracted from documents in the corpora, where $n$ will vary from one to some pre-determined maximum.  The example below demonstrates how text is converted to $n$-grams of various sizes. Here the $n$-grams are generated for the sequence ``beginning'', which has a maximum $n$-gram size of nine, and can itself be reduced to lower-order $n$-grams by reducing the sequence a character at a time, as illustrated in the table below:\\
\\
\begin{tabular}{ l  p{4cm}}
n$-$gram order & n$-$gram\\
\hline
9 & beginning\\
8 & eginning\\
7 & ginning\\
6 & inning\\
5 & nning\\
4 & ning\\
3 & ing\\
2 & ng\\
1 & g\\
\end{tabular}
\medskip
\\
For the collection model, $C$, the (global) probabilities of the character-based $n$-grams are stored in a suffix tree, while for the document model, $D$, the (local) probabilities are stored in a conventional database to make them easily available for use by other system components. From an implementation perspective each document is represented by a unique {\em document ID} and the positions of the $n$-grams in documents are stored with the probabilities inferred from the language model. Thus, when a user submits a query, Samtla will compute the probability that the query was generated by the model, $M$, where $M$ is either $C$ or $D$. In other words, Samtla will compute the probability that a user who is interested in a given document in the collection would submit that query. The documents in the collection can then be ranked according to the computed probabilities for these documents and the top scoring documents are returned to the user.\\
\\
We will now explain how the query model $P(q|M)$, denoted by $P_M(q)$ and read as ``the probability that the query $q$ was generated by the language model $M$'', is computed; we will assume throughout that $C$ represents the collection model. Using Bayes theorem \cite{Lewis1998}:
\begin{equation}\label{eq:query}
		P(M|q) = \frac{P_M(q)P(M)}{P(q)}
\end{equation}
\smallskip
The term $P(M|q)$ represents the conditional probability of the language model $M$ given the query $q$.
When $M$ is a document $D$, this is the probability of the document $D$ when the query is $q$, which will allow the system to rank the documents returned to the user. The right-hand side of (\ref{eq:query}) consists of the query model $P_M(q)$ multiplied by $P(M)$, the {\em prior} probability of the model $M$. When $M$ is a document model $D$, the prior is its presupposed probability, which is often assumed to be uniform, i.e. the same for all documents and can thus be ignored for the purpose of ranking (see Section~\ref{sec:conclusion} where we discuss a non-uniform prior); Finally, the denominator $P(q)$, i.e. the probability of the query, is the same for all documents and can thus also be ignored for the purpose of ranking. In summary, when the prior is uniform, we can rank documents according to $P_D(q)$, the query model for the document.\\
\\
Let $q = c_{1}, c_{2}, ..., c_{m}$ be a sequence of $m$ characters. Then, using the chain rule, the query model $P_M(q)$ is calculated as a product of conditional probabilities:\\
\begin{equation}\label{eq:chain}
P_M(q) = P_{M}(c_{1}, c_{2}, ..., c_{m}) = \prod\limits_{i=1}^{m}P_{M}(c_{i}|c_{1},..., c_{i-1}).
\end{equation}
\smallskip
Each conditional probability, $P_{M}(c_{i}|c_{1},..., c_{i-1})$, on the right-hand side of (\ref{eq:chain}), with $1 \le i \le m$, may be approximated by the maximum likelihood estimator (\emph{MLE}) \cite{Manning2008}:\\
\begin{equation}\label{eq:mle}
{MLE}_M(c_{i}|c_{i-n+1},..., c_{i-1}) = \frac {\#(c_{i-n+1}, ..., c_{i})} {\#(c_{i-n+1},..., c_{i-1})},
\end{equation}
where the \# symbol before a sequence indicates its raw count in the model $M$. Moreover, for any sequence of characters we only make use of its $n$ character history/context (or less than $n$ for shorter sequences) as an approximation to the conditional probabilities in (\ref{eq:chain}), in accordance with an $n$-order Markov rule \cite{Meyn2009}.
(We also take ${MLE}_M(c_1|c_0)$ to be ${MLE}_M(c_1)$.)\\
\\
A known problem with the maximum likelihood estimator (\ref{eq:mle}) is when the raw count of a sequence is zero, resulting in estimating the query model probability, $P_M(q)$ in (\ref{eq:query}), also as being zero. This may be the result of the user entering a character or word incorrectly, for example spelling mistakes or typographical errors.  Alternatively, the corpus may not be sufficiently large to encapsulate the full vocabulary of a given language, and thus due to this sparseness problem the query probability would be zero. To overcome this problem, \emph{smoothing} \cite{chengoodman1999, Zhai2004} adjusts the MLE probabilities to make them non-zero.\\
\\
We smooth the MLE probability (\ref{eq:mle}) via interpolation, using a weighted term, which defines the contribution to the overall probability for each order, $k$, where $k$ varies from a zero order, $0$-gram model, when $k=n+1$, to an $n$-gram model, when $k=1$. Each weight, represented by $\lambda_k$, defines the amount of interpolation, with lower-order models contributing less to the final probability.\\
\\
Thus our approximation of the conditional probabilities on the right-hand side of (\ref{eq:chain}) is given by the interpolation,\\
\smallskip
\begin{equation}\label{eq:approx}
\hat{P}_M(c_i|c_1,\ldots, c_{i-1}) \approx \sum_{k=1}^{n+1} \lambda_k {MLE}_M(c_i|c_{i-n+k},\ldots,c_{i-1}),
\end{equation}
\smallskip
where we use $\hat{P}$ to make clear that we are approximating $P$, and the weighted term for each $k$ is given by\\
\smallskip
\begin{equation}\label{eq:lambda}
\lambda_k =  \frac{n+2-k}{(n+1)(n+2)/2},
\end{equation}
\smallskip
where $n$ is the order of $n$-gram, which is composed by interpolating the $n$th order model with lower order ones. When $k=n+1$, then ${MLE}_M(c_i|c_{i+1})$ is taken to be the $0$-gram model, $\frac{1}{|V|}$, where $V$ is the finite alphabet of the language (for English this is 26 representing the characters of the Roman alphabet).\\
\\
If an $n$-gram of the query is not present for a particular document, then we need to back-off to a lower order $n$-gram. However, the $MLE$ score for the lower order $n$-gram will be too high because lower order $n$-grams are often more frequent in a document than higher order $n$-grams. To reduce the influence of a missing $n$-gram on the final query score, we back-off to the lower order $n$-gram, and as before, extract the probability for the back-off $n$-gram given by the document model $D$ and the collection model $C$ (obtained by revisiting the suffix tree). Assuming we obtain a match for the back-off $n$-gram, we smooth the probability with a weighted normalisation term $\frac{n}{n+2}$ to provide a consistent normalisation for each order of $n$ which defines the length of the $n$-gram obtained from the back-off procedure. If there is still no match, we store the result of $\frac{n}{n+2}$ and repeat the back-off process until we obtain a match for the lower order $n$-gram, or until we eventually arrive at the $0$-gram model. The smoothed probability for the missing $n$-gram is then the sum of the probabilities obtained from the back-off and is used to approximate the conditional probability $\hat{P}_M(c_i|c_1,\ldots, c_{i-1})$ in Equation~\ref{eq:approx}.\\
\\
The next stage is to smooth the conditional probabilities. The historians we are working with tend to submit long and verbose queries representing ``formulae'', which can impact on retrieval performance \cite{HustonCroft2010}, since they contain many uninformative query terms (such as prepositions: \emph{of, in, to, by}, and determiners: \emph{a, the}). To compensate for this we adopt the Jelinek-Mercer smoothing method, which involves a linear interpolation of the document model $D$ with the collection model $C$ using a coefficient represented by $\lambda$ to control the influence of each model on the final query score. \cite{Zhai2001}. The final smoothed query score for a document is obtained by replacing $\hat{P}$ in (\ref{eq:chain}) by $P$, as follows:\\
\begin{equation}\label{eq:smooth}
P_{D}(q) \approx \lambda \hat{P}_{D}(q) + (1-\lambda) \hat{P}_{C}(q),
\end{equation}
\smallskip
where we chose $\lambda = 0.6$, i.e. $60\%$ contribution from the document model to the smoothed query score. It is possible to further tune the value of $\lambda$ by experimentation. As mentioned, long verbose queries require more smoothing than keyword or title queries due to the number of uninformative terms and so may require a higher setting for $\lambda$ \cite{Zhai2001}. The further smoothing in (\ref{eq:smooth}) makes sense as even after the initial interpolation, the maximum likelihood document model probabilities may be low, while the maximum likelihood collection model probabilities will provide a better global estimate of the probability.\\
\\
To summarise, the list of documents $D$ containing the query are ranked according to $P_D(q)$ corresponding to the approximation in (\ref{eq:smooth}), which interpolates each document according to its language model and, in addition, interpolates the document and collection models. The assumption is that scoring documents in this manner presents to users the documents that are most likely to represent their information need.  We further emphasise that the Jelinek-Mercer smoothing method has been adopted, since many of our users will submit long queries representing textual fragments (for example, a Bible verse), and this method of smoothing has been shown to be particularly effective for addressing long and verbose queries \cite{Zhai2001, Zhai2004}. In the next subsection we will describe the suffix tree, which provides the lower level implementation of Samtla's language model.
\subsection{Suffix Tree}
\label{sec:suffixtree}
\begin{figure}[!tbh]
	\centering
    \includegraphics[scale=.5]{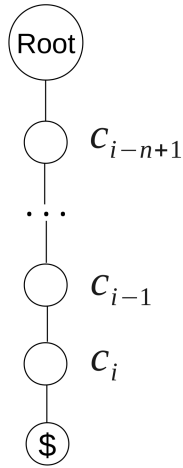}
	\caption{Calculating the MLE}
	\label{fig:suffixtreeMLE}
\end{figure}

Samtla's search capability, based on SLMs as described in the previous subsection, is supported by a space optimised character-based suffix tree, with the aim of holding the complete data structure in memory for fast retrieval \cite{Gusfield1997}.  \\
\\
In order to reduce the suffix tree's memory consumption, we create a $k$-truncated suffix tree \cite{Schulz2008}, which compresses the suffix tree by limiting its depth to $k$ nodes at most, and store the data attached to tree nodes in an external key-value database.  We have found that $k=15$ works well for the languages we have experimented with, after plotting the length of words present in the corpus, which showed that the majority were no longer than 15 characters in length.\\
\begin{figure}[!tbh]
	\centering
		\includegraphics[scale=.5]{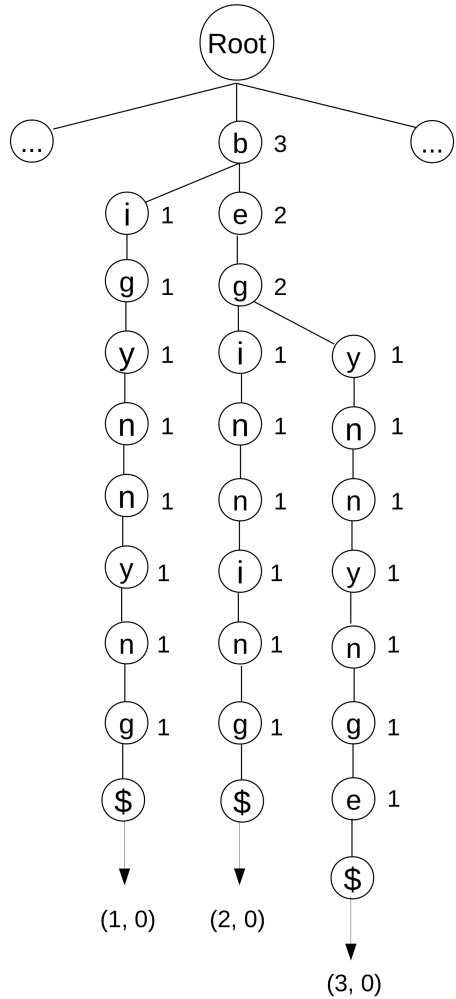}
    	\caption{A truncated suffix tree over the strings ``beginning', ``bigynnyng', and ``begynnynge''}
	    \label{fig:suffixtree}
\end{figure}
\\
A further method to reduce the space requirement of the suffix tree is to compress dangling nodes, which are nodes that have only a single descendent (or child). These are gathered together during a depth-first-search and stored as a 'supernode', whose label is constructed from the concatenation of the collected node labels \cite{Gusfield1997}.\\
\begin{figure}[!tbh]
	\centering
		\includegraphics[scale=.5]{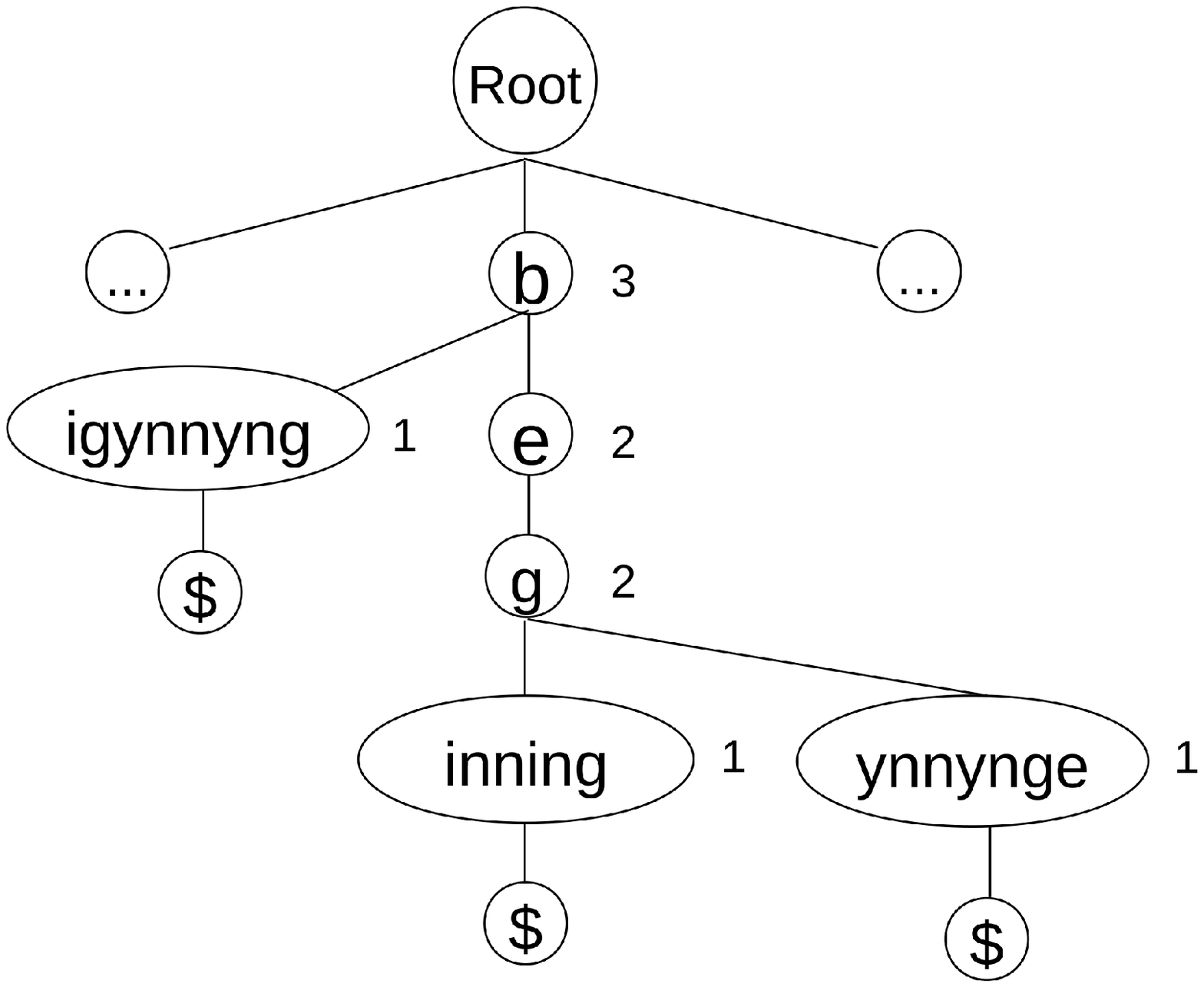}
	    \caption{A compressed suffix tree where the super-nodes are rendered as ellipses}
	    \label{fig:suffix_tree_comp}
\end{figure}
\\
Given a text string, the resulting suffix tree represents a compressed ``trie'' data structure containing all the suffixes of the string as their keys and positions in the string as their values. A generalized suffix tree is a suffix tree constructed from a set of text strings instead of a single one, so its leaf nodes (represented by \emph{\$} in Figure~\ref{fig:suffixtree}) only need to store the ID and start position of the character string.  In each node of the generalised suffix tree constructed from the corpus (document collection), we store the frequency of the corresponding string of characters, which is later used to perform the above-mentioned maximum likelihood estimations for the character-based $n$-gram SLM. \\
\\
Once the generalised suffix tree is constructed, we calculate the conditional probabilities which form the basis for the collection model $C$ (discussed in the previous section).  The tree is traversed starting at the root node.  For each node, the conditional probability is calculated by dividing the count of the current node with the count of its parent node, as defined in Equation~\ref{eq:mle} above, and illustrated in Figure~\ref{fig:suffixtreeMLE}.\\
\\
After calculating the initial probabilities the tree is traversed a second time to apply the interpolation between each order of $n$-gram (as detailed in Equation~\ref{eq:approx}) to produce the final smoothed collection model $C$.\\
\\
Searching a suffix tree is performed by starting at the root node and then descending the tree along a unique path by comparing characters of the query with the label stored at each node. When the characters of the query are exhausted or a mismatch occurs, the sub-tree rooted at the last-matched node is traversed with a breadth-first traversal, and all leaf nodes are collected resulting in an index of document IDs and start positions. Partial matches are also obtained during the traversal as we are always returning the last matched node, regardless of whether there are any further characters to match. (although see Section~\ref{sec:Relatedqueries} for a discussion on how we can trigger different search strategies when there is a mismatch for the full query)\\
\\
The language agnostic nature of the implementation has been tested with a number of corpora, which are both different in terms of structure, but also in terms of the language, dialect, script, and domain.
\section{Text Mining Tools}
\label{sec:textmining}
Books, web pages, articles, and reports are all examples of unstructured text data where relevant information exists potentially anywhere within the document.  Unstructured text data is often managed and retrieved via a search engine (see \cite{Levene2010}). Search engines provide the means to retrieve information but not to analyse it, this is where text mining techniques are useful, as they provide the researcher with different views of the data that can enable them to discover and evaluate textual patterns \cite{Aggarwal2012}.\\
\\
The Samtla system is designed as a research environment packaged with a set of extensible text mining tools. The tools provide a means to analyse a corpus through the identification of patterns or ``formulae'' that are of potential interest to the researchers.  In addition, the tools have been developed alongside each user group in order to identify the problem domain and provide solutions which can be implemented in accordance with the probabilistic approach adopted by the underlying system. \\
\\
Each tool is built on one or more components of the data model (illustrated in Figure~\ref{fig:architecture} and discussed in Section~\ref{sec:datamodel}).  For example, Samtla uses the collection model $C$ and the suffix tree data structure to provide a related query tool, which generates and ranks queries similar to the users original search term (see Section~\ref{sec:Relatedqueries}).  The related documents feature measures the similarity between document pairs using the language model $D$ for each document, which is then ranked and presented to the user as a list of documents similar in content to the document they are viewing (see Section~\ref{sec:Relateddocuments}). This subset of tools fall under the recommendation component of the system, where statistical language models have been shown to perform well \cite{Lavrenko2000}. \\
\\
The tools are context-dependent as the system only presents the user with the results from a given tool if it makes sense in the given context.  For example, if a user is viewing a document (document view), Samtla displays an informational side-bar containing related documents, selecting a document from the related documents list directs the user to an interactive document comparison tool (see Section~\ref{sec:doccomp}) where shared-sequences can be compared between the two documents.  Whilst in document view, the user has access to metadata for each document, which is tailored to each user group. In document view the user can also overlay additional data in the form of {\em named entities} \cite{Nadeau2007}, which are labelled with additional metadata from external sources such as Wikipedia,  online encyclopedias, and the Google maps API. The context-dependent design of the tools results in a minimal user interface, which dedicates more screen real estate to the data and tool output.
\subsection{Related Queries Tool}
\label{sec:Relatedqueries}
The related queries tool extracts 10 text fragments from the corpus (specifically, from the suffix tree representing the collection model $C$ as discussed in Section~\ref{sec:LM}), that are most similar to the user's original query.  These are then displayed as part of the ranked search results.  For example, searching for ``beginning'' in the Bible edition of Samtla would return related queries ``bigynnyng'' and ``begynnynge'', which represent alternative spellings. The related queries are generated through a process similar to the Levenshtein edit distance algorithm \cite{Gusfield1997} where alternative forms of the original query are created through processes of deletion, substitution, and insertion. These can be defined more formally as follows. Let $Q$ represent the related query, where $n$ is the length of the original query $q$, and $i \in {1, 2, \ldots, n}$. Then the string edit processes are defined as:\\
\\
\begin{tabular}{ p{2.5cm}  p{8cm} }
\textbf{Method} & \textbf{Related Queries} \\
\hline
Deletion & $Q = q_{1}, q_{2}, ..., q_{i-1}, q_{i + 1}, ..., q_{n}$\\
Substitution & $Q = q_{1}, q_{2}, ..., q_{i-1}, $? $, q_{i + 1}, ..., q_{n}$\\
Insertion & $Q = q_{1}, q_{2}, ...,q_{i-1}, $? $, q_{i}, ..., q_{n}$\\
\hline
\end{tabular}\\
\\
\linebreak
As an example, if the original query is ``beginning'', then the following related queries are generated.\\
\\
\begin{tabular}{ p{2.5cm}  p{8cm} }
\textbf{Method} & \textbf{Related Queries} \\
\hline
  Deletion & eginning, bginning, ..., beginnig, beginnin \\
  Substitution & ?eginning, b?ginning, ..., beginni?g, beginnin? \\
  Insertion & ?beginning, b?eginning, ..., beginnin?g, beginning? \\
\hline  
\end{tabular}\\
\\
\linebreak
When the user submits a query, related queries are automatically extracted from the suffix tree component of the system.  This is achieved by replacing each character of the original query one at a time with a wild-card character and then submitting them to the tree.  As the wild-card character is not indexed by the suffix tree there will be a guaranteed mismatch at that point in the query.  When a mismatch occurs, we execute the above functions, which traverse the suffix tree from the mismatched node, and attempt to match the remainder of the query. Deletion does not require a wild-card character since we simply remove a character from the string where the wild-card character would appear. Insertion and Substitution are achieved by replacing the wild-card character with the node label of each child rooted at the last matched node (its parent node), and attempt to match the remainder of the query after the wild-card character. All successfully matched strings are returned as a ranked list of potential related queries according to their smoothed probability scores from the collection model $C$.\\
\\
As illustrated above, the combination of the suffix tree data structure and the collection model component $C$ of the language model provides a good basis for constructing a query recommendation system. 
\subsection{Related Documents Tool}
\label{sec:Relateddocuments}
Related documents are those documents that have common string sequences shared between them. The related documents tool finds up to twenty documents from the corpus that are most similar to the document currently viewed by the user (referred to as the target document).  The retrieval of twenty documents is a user interface decision, since we compute similarity scores over all document pairs in the corpus. Each document, in the related documents menu, represents a link to the document comparison tool (discussed in Section~\ref{sec:doccomp}).\\
\\
A document can be considered a probability distribution over $n$-gram sequences \cite{Endres2003}, and the similarity between a pair of documents may be calculated through the Jensen-Shannon Divergence(JSD) over their corresponding document models $M_d$ (as described in section~\ref{sec:LM}) \cite{Lin1991}. The JSD is the symmetric version of the well-known Kullback-Liebler Divergence (KLD) defined as:\\
\[ D_{KL}(P||Q) = \sum_{i} P(i) \log_{2} \frac{P(i)}{Q(i)} \ ,\]\\
where in our context $P$ and $Q$ represent two smoothed $n$-gram probability distributions provided by the corresponding document models, and $i$ is a value drawn from the respective smoothed $n$-gram distribution based on a sliding window of size $n$.  The smaller the sliding-window size $n$, the finer-grained the document similarity measure. In our experiments we have found that $n=7$ provided a good balance based on our 15-gram language model. The JSD is calculated as\\
\[ JSD(P||Q) = 1-\sqrt{\frac{1}{2} D_{KL}(P||M) + \frac{1}{2} D_{KL}(Q||M)} \ ,\]\\
where $M$ is the average of the two distributions $\frac{1}{2} (P+Q)$ \cite{Endres2003}. The resulting $JSD$ produces a score between 0 and 1, where a score of 1 means the documents are identical. The $JSD$ scores are ordered in descending-order according to their similarity to the target document so that the most similar documents are ranked at the top of the related documents list.
\subsection{Document Comparison Tool}
\label{sec:doccomp}
A common task applied to language corpora is to find representative examples of language use exemplified by string patterns.  Samtla provides a document comparison tool, where users can compare the document they are currently viewing with a document selected from a list of similar documents (discussed above in Section~\ref{sec:Relateddocuments}). This feature is considered to be essential by our user groups, since this form of comparison was performed manually and can be a complex task due to overlapping shared-sequences.   \\
\\
Although there exist some document comparison tools such as \texttt{diff} on UNIX, they perform \emph{global sequence alignment}, which attempts to match the entire documents, while Samtla users are interested in \emph{local sequence alignment} which identifies text regions of similarity within two documents that could be widely divergent overall. The underlying algorithm for identifying shared text patterns is a tailored variant of the Basic Local Alignment Search Tool (BLAST) algorithm \cite{Ma2011}, widely used in bioinformatics for comparing DNA sequences. \\
\\
We first extract all trigram character strings shared by the given pair of documents as seeds, and then extend those seed strings one character at a time, first from the left and then from the right. During the iterative extension process, we score any pair of (approximately) matched strings $s_{1}$ and $s_{2}$ by their Levenshtein edit distance \cite{Gusfield1997}, which measures the number of changes required to convert one string in to another using deletion, insertion, and substitution. The metric is defined as:
\begin{equation}
\label{eq:eq16}
ed(s1, s2) \leq \lfloor m \delta \rfloor
\end{equation}
\\
The extension stops when the edit distance reaches the floor of a certain threshold $m \delta$, where $m$ is the length of the shorter matched string between $s_{1}$ and $s_{2}$,  multiplied by a tunable tolerance parameter $\delta$. The default setting is $\delta=0.2$ which is equivalent to a 20\% difference between the two sequences, before moving on to the next seed.  Characters representing punctuation are ignored during the extension process.\\
\\
As an example, a text pattern found by Samtla in the Bible corpora, starting from the trigram seed string \textit{ham}, is shown as follows.\\
\\
\textsc{King James Bible}: \underline{Noah; Shem, \textbf{Ham}, and Japheth}\\
\textsc{Douay-Rheims Bible}: \underline{Noe: Sem, C\textbf{ham}, and Japheth}\\
\\
The above example has a total edit distance of 4. The strings \emph{Noah} and \emph{Noe} have an edit distance of 2, since there is one substitution (\emph{a} $\rightarrow$ \emph{e}) and one deletion (final character \emph{h}) required to convert \emph{Noah} to \emph{Noe}. The strings \emph{Shem} and \emph{Sem} require one deletion of character \emph{H} which is equal to an edit distance of 1, and likewise, \emph{Ham} and \emph{Cham} is converted with the insertion of character \emph{C} at the beginning of the string.\\
\\
It can be seen that our method captures text patterns, which differ in terms of spelling errors or orthographic variations. Despite such superficial differences, a researcher of Bible scripture would probably consider those two text fragments as identical passages (Chapter 10, Genesis). For a discussion of the document comparison interface see Section~\ref{sec:comparedocs}.
\subsection{Named Entity Tool}
\label{sec:NER}
Named Entity Recognition (NER) \cite{Bikel1999} describes the process of extracting words (or sequences of characters, in our case), that represent names of people, companies, and locations.  Samtla uses gazetteers to extract named entities from the raw documents. Gazetteers have been used for some time to improve the performance of named entity systems, other more sophisticated methods exist, for instance, semi-supervised learning techniques such as bootstrapping \cite{Aggarwal2012}, however gazetteers are becoming popular once again due to the wealth of structured data on named entities provided by platforms such as Wikipedia and DBpedia \cite{KaTo07}.  A further motivation for adopting the gazetteer approach, is that the current versions of Samtla support a number of historic text collections, such as the Bible and Vasari's 'the lives of the most excellent artists and architects' \ref{sec:casestudy}, these collections represent closed corpora, which means that there are rarely going to be new documents added to the collection. Consequently, gazetteers are sufficient for these types of domain specific and static corpora as there are a wealth of lists already compiled by researchers that can be used to form the basis for gazetteers. Furthermore, we have found that Wikipedia can be leveraged since there are a large number of general lists of people, locations, and other miscellanea \cite{wikilists}, but also lists for specific collections \cite{biblicalnames, biblicalplaces}.\\

Named entities are located in the documents by submitting each entry in the gazetteer to the suffix tree as queries. Each full match is stored in a database organised by entity type together with the document identifier and an index of start positions in the text.  The data is parsed to the browsing tool (see Section~\ref{sec:browsedocs}), which provides further entry points to the documents, with the named entities themselves being rendered as an additional layer over the document in document view (further discussed in Section~\ref{sec:UI}).\\

The gazatteers could also be used to form the basis of training data for a statistical learning approach \cite{Fayyad1996} to enable Samtla to identify and mark up documents semi-automatically, which is an approach that we will be investigating as part of future work.

\subsection{Recommendation Tools}
\label{sec:community}
Personalised recommendation systems are familiar to many users of the internet. For instance, online shoppers often encounter the `what other customers bought' page, which presents a series of recommended items that other buyers purchased based on items in their 'basket' or 'shopping cart'.  Other examples of recommendation systems can be found in socially affected personalisation where a user is part of a select group who share content and opinions with other users they trust, as well as collaborative search, which enables users to discover new search terms based on the search behaviours of other users \cite{Wilson2011}.  Samtla leverages user activity to generate recommended queries and documents.  Through analysing the log data, Samtla can inform users of the top-10 most popular queries and documents in the research community, so as to support users' collaborative search. Thus a user of Samtla can be directed to the ``interesting'' aspects of the corpus being studied, which may not have occurred to them previously.  \\
\\
Log files are used to store usage statistics, user interaction through the system using referrer URLs, and system error reports.  This data can be leveraged in interesting ways, one of which is to return the users search and page view history. Users may also wish to discover what is popular in a corpus, as a way to find new documents of potential interest.  The current version of Samtla supports a community feature which suggests search terms and document views based on their popularity, this requires storing data such as unique userIDs, timestamps, queries and document IDs.  The user data is then used to produce top-ten ranked lists of queries and document views per user and the community as a whole.\\
\\
The popular queries and documents are ranked and selected using an algorithm similar to the Adaptive Replacement Cache (ARC) \cite{Megiddo2004}, where the frequency of each query or document is combined with its recency (measured by the number of days that have passed since the last submission of the query or document view), and used for ranking. This ensures that the recommended popular queries and documents are biased towards fresh ones and updated along with time.  Formally the popularity of a query or document is defined as\\
\begin{equation}
		 popularity = T^{\beta} R^{1-\beta}
\end{equation}
\\
where \emph{T} = $\frac{1}{S}$ with $S$ representing the count in days since the last submission, where $today$ = 1, with weighted term $\beta=0.6$, and parameter \emph{R} which represents the raw count of submissions for the query or document.  The combination of the two terms $T$ and $R$ prevent submissions with high counts, but longer time between submissions, from dominating the top entries of the recommended queries or documents. \\
\\
The resulting ranked results are made accessible via the respective side-bars in the user interface, which are populated when the user navigates to a document through browsing or searching (see Section~\ref{sec:NER}). 
\section{User Interface}
\label{sec:UI}
Interaction with the system is through a browser-based interface, the user can perform three main tasks in the current version of the system: (i) corpus browsing (see Section~\ref{sec:browsedocs}), (ii) search (Section~\ref{sec:searchdocs}), and (iii) document comparison (Section~\ref{sec:comparedocs}).
\begin{figure}[!tbh]
	\centering
		\fbox{\includegraphics[width=\columnwidth]{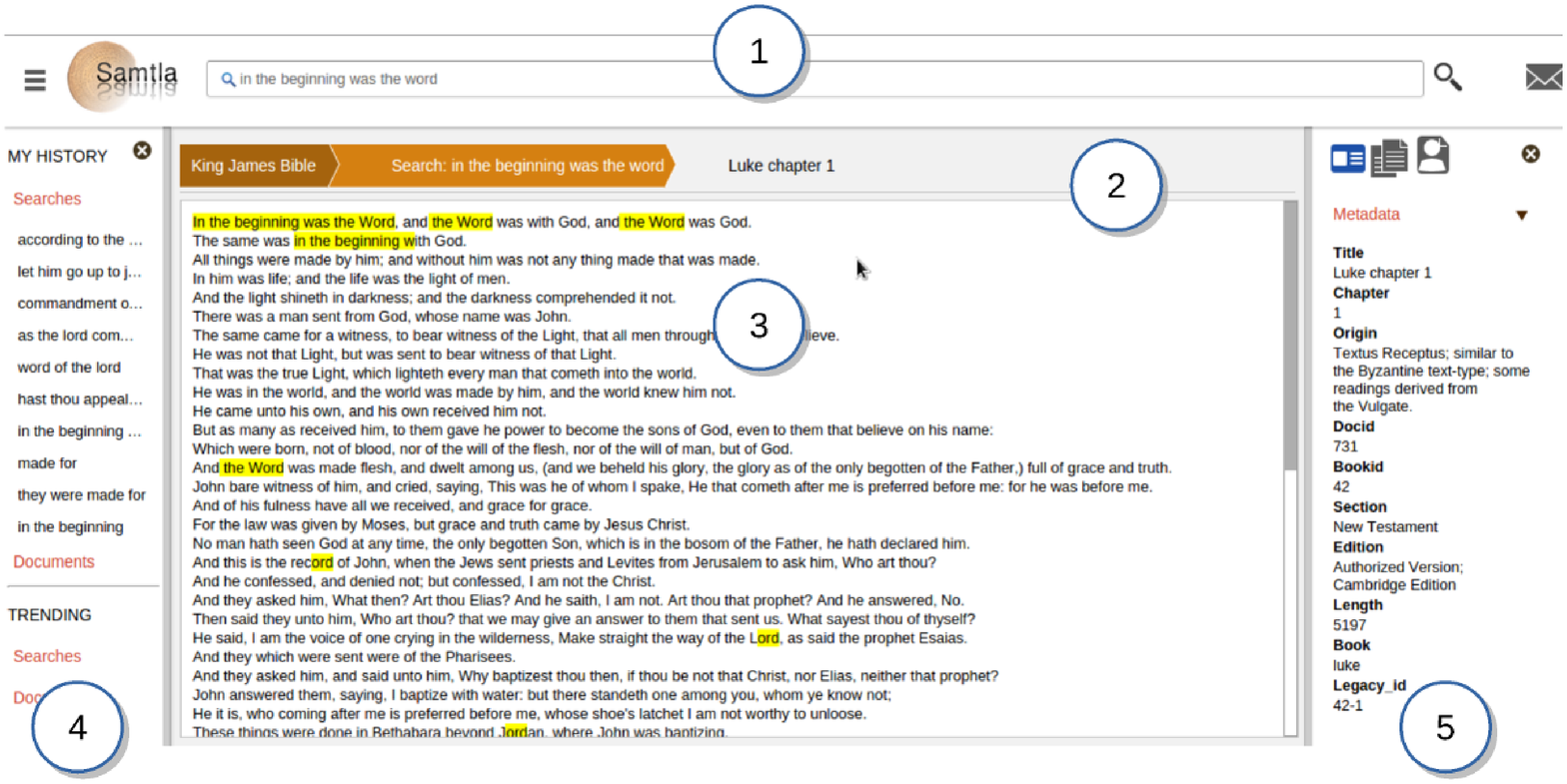}}
	\caption{Samtla User Interface showing 1.) the search bar, 2.) breadcrumb-based navigation 3.) main window, which displays search results, and the document text with additional information such as highlighted query terms (shown here), and output of the text mining tools, 4.) the query and document view history for the user and community (trending searches and documents), and 5.) a side-panel for displaying the metadata, related documents (for accessing the document comparison tool), and activating additional data layers over the document e.g. named entities.}
	\label{fig:UIstructure}
\end{figure}\\
The tool set is designed to be modular and extensible in order to enable further tools to be developed with our users without affecting previously established system components (see Sections~\ref{sec:architecture}, ~\ref{sec:casestudy}, and \ref{sec:conclusion}).
\subsection{Browse Documents}
\label{sec:browsedocs}
\begin{figure}[!tbh]
	\centering
		\fbox{\includegraphics[width=\columnwidth]{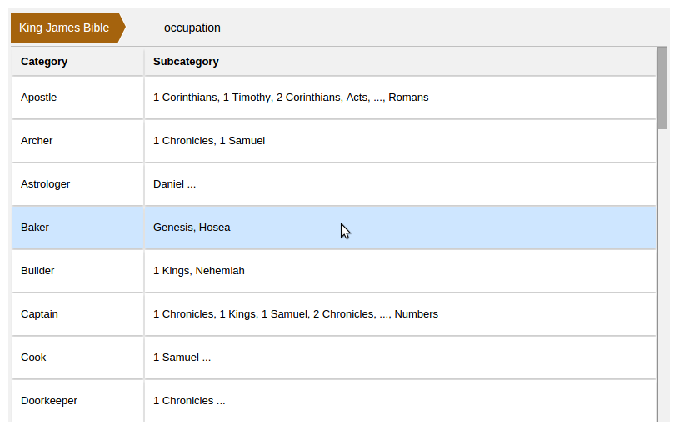}}
	\caption{Browsing the Bible corpus using the list view}
	\label{fig:listview}
\end{figure}

\begin{figure}[!tbh]
	\centering
		\fbox{\includegraphics[width=\columnwidth]{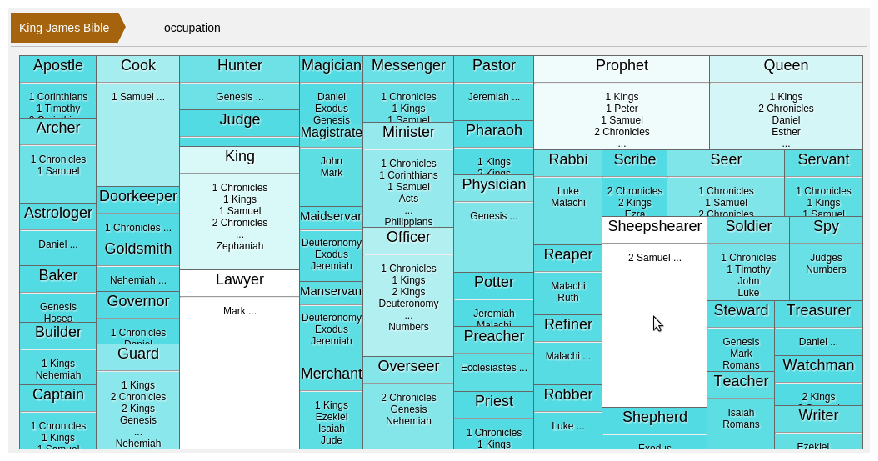}}
	\caption{Browsing the Bible corpus through the treemap view generated from the document metadata}
	\label{fig:treemap}
\end{figure}

Samtla adopts a clustered (or faceted) navigation model, where each cluster describes a category represented by a collection of documents sharing a common property \cite{Broughton2002} \cite{Crain2012}.  Clustering documents according to a particular feature \cite{Steinbach00}, can provide users with an indication of the type and availability of data in a system \cite{hearst2009}, and is a useful approach for encouraging users to explore and discover information within a collection \cite{hearst2008}. By adopting a clustered navigation model, future components can be integrated in a modular fashion, without introducing visual clutter through traditional UI elements like tabs and drop-down menus. \\
\\
The browsing architecture is divided in to two separate presentation layers.  The default is a list view, which mimics a traditional file directory where each row entry represents either a folder or an individual document. Columns contain the cluster label or document name, and further information extracted from the document metadata (see Figure~\ref{fig:listview}).
The alternative view uses a squarified {\em treemap} \cite{Bruls99squarifiedtreemaps}, and can be considered as representing a topic model (see Figure~\ref{fig:treemap}), where each topic provides the user with a different clustered view of the corpus generated from the metadata or the named entity tool (see Section~\ref{sec:NER}).  The user can switch between the list and treemap views via a button in the interface, as users may prefer one form of presentation over another.\\
The advantage of the treemap representation is that it is very flexible and can be enriched with textual or visual information by populating the cells with metadata or by altering the dimensions or colour of the cells to indicate membership or extent. For instance, the size of each cell can be adjusted to reflect that a document is longer or that a cluster contains more members than others or a preview image of the document could be displayed to aid navigation.
\pagebreak
\subsection{Search Documents}
\label{sec:searchdocs}
\begin{figure}[!tbh]
	\centering
		\fbox{\includegraphics[width=\columnwidth]{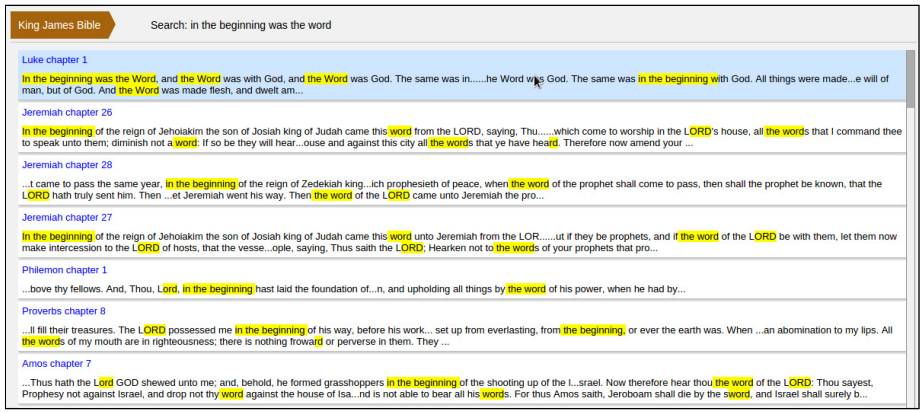}}
	\caption{The search result interface - showing ranked documents with snippet windows}
	\label{fig:search}
\end{figure}

Query results in Samtla are divided in to two types: exact and partial matches. Partial matches do not encompass the full query, in other words, not all characters of the original query were matched.  However, partial matches are returned, since they could still be of interest to the researcher, but will appear lower down in the search results below full query matches. For example, if we consider the query ``\emph{pharaoh was wroth against his two officers against the chief of the butlers}'', from the Bible, Samtla will return (aside from exact matches), examples of other roles exemplified by the inexact matches ``\emph{the chief of}'', such as; ``\emph{chief of the cup-bearers}'', ``\emph{chief of the bakers}'', ``\emph{chief of the tower/round-house}'', and ``\emph{chief of the eunuchs}'', which tells us something about roles within the King's court at that time (see Figure~\ref{fig:search}).  Alternatively, the results returned by a partial match may help the researcher to reformulate their query given them a basis to start from - in contrast to boolean forms of search, which return no results if the exact query was not found.  \\
\\
When a search is performed, we obtain an index from the suffix tree (see Section~\ref{sec:suffixtree}) containing the document id and the start and end positions of the matched $n$-grams for each document.  This is passed to the language model for ranking the documents according to the query, and a snippet generation tool for producing short snippets of the match query for the search result view. We order the ranked list by sorting the results first according to the length of the matched query, and then by the probability of the query given the query score for the document, retrieved from the SLM (see Section~\ref{sec:datamodel}).  The top ranks of the results reflect the highest scoring or most relevant documents. \\
\\
Each document in the search results is then rendered with its title and the generated snippet window showing the preview of the query in the document, which contains the top-3 snippets that best describe the query. Snippet windows were selected as the most appropriate method for summarising the document, as they are familiar to users \cite{hearst2009}.\\
\\
The snippets enable the user to evaluate the relevance of each document in the ranked list before deciding which document best meets their information need. Snippet length is tunable and we define a parameter $w$, which limits the maximum length of each snippet, for our purposes this is set to $w=100$ characters, however in future versions this could be provided as a user setting.  \\
\\
The snippet scoring algorithm extracts all potential snippets and ranks them by interpolating the length of the set of ngrams in the query found in the snippet with the total count of all terms appearing in the snippet, where more weight is assigned to snippets containing all of the query terms. This ensures that the snippet window is ranked in such a way that the top snippets will contain all of the terms of the users' query before presenting snippets with only partial matches of the query. Let $\delta$ be the cardinality of the set of $n$-grams that are present in the snippet, which are present in the query and $\mu$ be the count of all query $n$-grams (including repetition) found in the snippet, with a weighted term $\alpha = 0.9$, which as mentioned, biases the snippet towards those that contain all parts of the query, then each snippet can be scored through,  
\begin{equation}
    snippet\ score = \delta^{\alpha} \mu^{(1 - \alpha)}, 
    \label{eq:eq15}
\end{equation}
\\
The snippets are then sorted in descending order by the score returned in Equation~\ref{eq:eq15}, and the top-3 selected as a preview for the document.  When the user selects a snippet the system opens the document and scrolls to the location of the selected snippet.  Other components of the search interface include the related queries, which are displayed above the search results as links ordered by their probability given the collection model $C$ (see Section~\ref{sec:datamodel}), from left-to-right.
\subsection{View Documents}
\label{sec:viewdocuments}
\begin{figure}[!tbh]
	\centering
		\fbox{\includegraphics[width=\columnwidth]{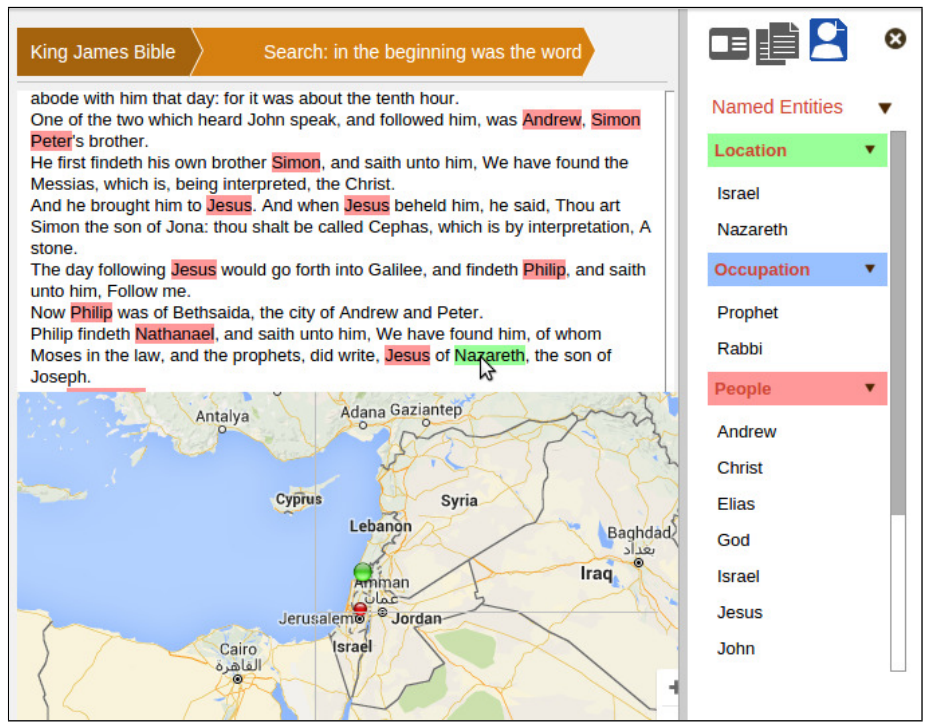}}
	\caption{The Bible version of Samtla showing the document view with the named entity layer.}
	\label{fig:named_entity_bible}
\end{figure}

When a user arrives at the document level through browsing or searching, they are presented with a main window displaying the document text, or where available the image of the scanned document, see Figure~\ref{fig:UIstructure}. If the user has navigated to the document through the browsing tool, then any metadata related to the document, including the named entity (see Figure~\ref{fig:named_entity_bible}, is highlighted, for documents located using the search tool, we highlight all instances of the query. In document view, the user has access to the metadata, document comparison, and named entity tools.   

\pagebreak
\subsection{Compare Documents}
\label{sec:comparedocs}
Related documents (see Section~\ref{sec:Relateddocuments}), provide access to the document comparison tool. The document comparison tool is composed of two document windows, one for the target document (the document the user is currently viewing), and another for the document selected from the list of related documents.  Each time a user selects a new related document, both documents are updated with new sequence data and the longest shared-sequence is highlighted in each as a starting point for the user.
\begin{figure}[!tbh]
	\centering
		\includegraphics[width=\columnwidth]{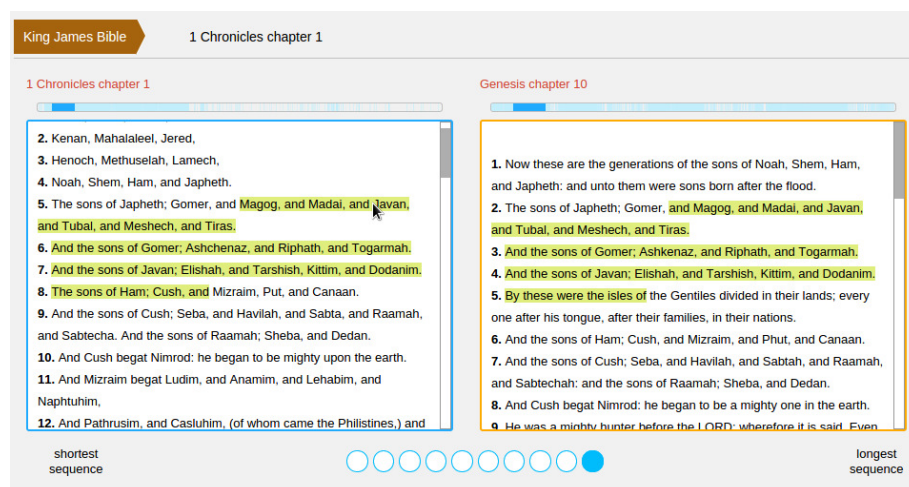}
	\caption{An example of Samtla document comparison. The document comparison interface shows a pairwise comparison of the target document (left) and a document selected from the list of related documents (right). Sequences highlighted in yellow reflect the currently selected sequence, and blue represents all sequences shared between the two documents.}
	\label{fig:doc_comp}
\end{figure}
The tool is equipped with a control to choose the length of the shared text pattern to view, with the minimum being 3-gram and the default setting displaying the longest sequence found between the two documents.  This enables users to investigate large shared-sequences spanning several lines to smaller sequences representing a word or grammatical affix (typically 3-gram in length).  Appearing above each document is a small horizontal map summarising all shared-sequences in the document, which provides the user with an overview of how the sequences are distributed throughout the two documents. For example, the shared-sequences may all appear in the introduction or abstract of the text. Clicking on a shared-sequence in a document highlights all instances of that sequence across both documents (see Figure~\ref{fig:doc_comp}).  
Sequence comparison is difficult to perform manually, especially over several documents and particularly when some of the sequences may be approximate, or overlap one another.  The design of the document comparison tool is based on feedback from our users, and the orientation of the document windows attempts to emulate the manual process of document comparison, where a user may layout two documents, or pages side-by-side.
\section{Case Studies}
\label{sec:casestudy}
\begin{figure}[!tbh]
	\centering
		\fbox{\includegraphics[width=\columnwidth]{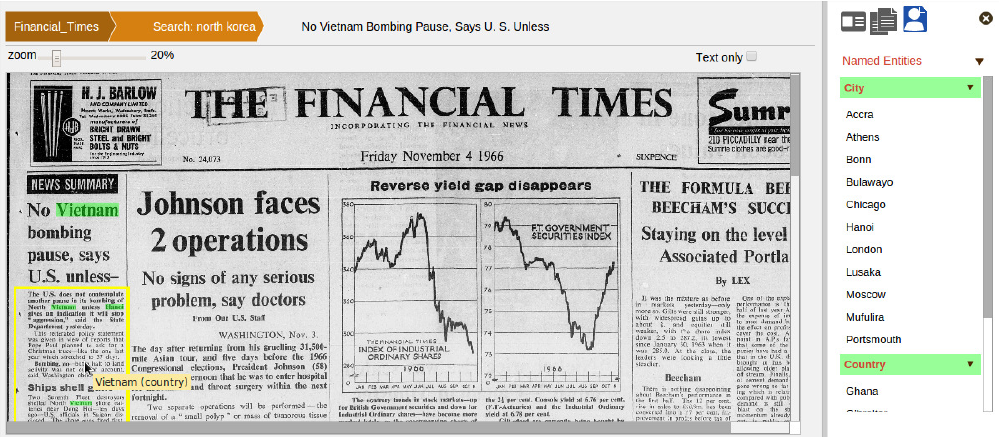}}
	\caption{The FT version, with the document view showing the named entity layer rendered on top of the original image.}
	\label{fig:FT}
\end{figure}

There are currently five versions of Samtla, with two versions serving two separate groups of digital humanities researchers. The first user group is represented by a team of historians led by the University of Southampton\cite{VMBA} who are analysing a corpus of 650 Aramaic Magic Bowls and Amulets from Late Antiquity (6th to 8th CE) written in a number of related dialects including Aramaic, Mandaic, and Syriac. The texts are written in ink on clay bowls, and cover a wide subject matter. The research involves searching and comparing textual fragments, which are formulaic in nature and provide an insight into the development of liturgical forms which differ due to transmission over centuries, and orthographic variation as a result of differences in authorship or dialect.  There are also transcription errors resulting from damage to the original artefact, or illegible characters.  Existing tools were not sufficient for identifying approximate text fragments meaning the analysis was largely a manual process of comparison and documentation.  \\
\\
The second user group is the Vasari Research Centre. The documents represent chapters from the book \emph{Lives of the Most Excellent Painters, Sculptors, and Architects} by Giorgio Vasari (1511 - 1574).  Giorgio Vasari is considered to be the founding father of the Art History discipline \cite{DAH}.  The Vasari Samtla contains documents in the original Italian and a corresponding English translation, and is used for research and as a teaching aid for students in class.  Users can view either version by searching, browsing, or selecting the alternative version from the metadata in document view, which is then displayed side-by-side for comparison (see Section~\ref{sec:UI} for more detail on the User Interface).  The Vasari corpus also contains a large number of images of paintings and architecture, which are displayed with the document metadata.\\ 
\\
A third version of Samtla is applied to the Microsoft corpus of 68,000 scanned documents, which was bequeathed to the British Library. The collection represents books digitised from the worlds libraries and contains a range of languages, literary genres, over a couple of centuries. Moreover, as a proof-of-concept, a special edition of Samtla has been applied to the King James Bible, in English.  This version is used for demonstration purposes and evaluation as many people are familiar with the content of the Bible. \\
\\
The most recent Samtla was constructed for a pilot study between the British Library and the Financial Times (FT). The documents are represented by a corpus of Newspapers that have been digitised using OCR technology.  The OCR data was provided along with the scanned pages of the Newspaper, which cover the year 1888, 1939, 1966, and 1991.  This particular archive, required new tools that could leverage the image data in order to compensate for poor quality OCR that reflected the current state of the art at the time of digitisation.  Much of the text for the earlier articles (e.g. 1888, 1939, and to some degree 1966) are not reliably searchable due to poor recognition rates, and consequently the focus was on developing a metadata search component to complement the existing search tool, allowing users to search both the metadata and the full document. In addition, this Samtla presents users with the original image (see Figure~\ref{fig:FT}), which utilises the document metadata to render boundaries around the articles and to make them selectable so that users can navigate the articles contained in a single newspaper.  \\
\\
The user is able to navigate between the raw OCR text and the scanned image, which required that some existing tools required some adaption to make use of the image data. For instance, the named entity tool (see Section~\ref{sec:NER}) was originally developed for text data, but the FT version renders the named entities in both the raw text and the original scanned image, providing the ability to select and filter named entities in both views.  \\
\\
Each version of Samtla differs only in terms of the respective document collection, while the underlying system remains unchanged due to its language-independent and data-driven design. Upgrades and new features are rolled out across all versions, meaning that all user groups benefit from tools developed in collaboration with each user group.
\section{Evaluation}
\label{sec:evaluation}
\subsection{Overview}
In this section we describe the evaluation process for measuring the performance of the Statistical Language Model underlying the Samtla search engine (see Section~\ref{sec:datamodel}). The evaluation assesses the ranking quality of the Samtla search engine in terms of whether the system (see Section~\ref{sec:datamodel}) consistently puts the most relevant documents at the top positions of the search results list. \\
\subsection{Crowdsourcing}
Crowdsourcing is a web-based business model \cite{Brabham2008} that enables companies and individuals to employ the skills of people from a distributed community in order to perform some task in return for a small reward.  These tasks are often large in scale or complex, and therefore time consuming as a result.\\
\\
Crowdsourcing in Information Retrieval has generally involved outsourcing manual tasks such as data-annotation, labeled-data collection for training models, and system evaluation. This process was often completed in-house with a limited workforce, which depending on the size of the task, could be a slow process involving several days of work.  Due to the size of the crowd, who are globally dispersed, tasks can be completed much faster at any hour of the day.  There is also the potential for reducing bais in aggregated results, compared to in-house evaluations, due to the diversity and representativeness of the workers in terms of demographic \cite{Lease2013}.\\
\\
There are a number of crowdsourcing platforms available for running surveys and evaluations. Amazon Mechanical Turk\footnote{\url{https://www.mturk.com/}} ({\em MTurk}) is one of the better known ones \cite{Kittur2008, Alonso2011, Zaidan2011}, however, it is only available to researchers resident in the United States of America.  As a result we selected Prolific Academic\footnote{\url{https://www.prolific.ac/}} \cite{ProlificAcademic}, a crowdsourcing platform for academics and part of the Software Incubator at the University of Oxford. Prolific Academic currently have a participant pool of over 22,000 participants (as of 27/12/2015).  The platform directs users to a website hosting a static survey or application.  When the user completes the survey, they are presented with a URL, which activates a payment for their completed submission.\\
\\
The majority of crowdsourcing platforms we investigated provided support only for static surveys, where the evaluation is represented by a series of static web pages constructed using a template web form editing tool. Platforms that allow the researcher to link to a URL hosting a web application, provide more flexibility by enabling the evaluation software to perform some action or logic based on user input, including monitoring the quality of the results, or distributing groups of tasks across different groups of users.  Like MTurk, Prolific Academic directs users to an external website hosted by the researcher, which makes it possible to implement a survey tool that can serve dynamic content to the users, and also record information in the form of log files during the evaluation.
\subsection{Methodology}
The evaluation consisted of 50 queries represented by a ranked list of the top-10 documents returned by the Samtla system.  The users were asked to assign graded relevance scores according to the four relevance grades ``not relevant'', ``somewhat relevant'', ``quite relevant'', or ``highly relevant'' to each document in the ranked list based on a given query displayed at the top of each search result.
\subsubsection{Data Preparation}
To evaluate the ranking performance of the system we used the King James Bible version of Samtla, since many people are familiar with the content of the Bible to some degree.  We prepared a set of 50 queries of variable length, ranging from single word queries (i.e. ``Moses'', and ``Jesus Christ''), to longer verbose queries representing set phrases (i.e. ``the Lord hath spoken'', and ``blessed be the Lord'').  We also constructed two test queries in order to have some control over the quality of users.\\
\\
Each query was submitted to the Samtla search engine and the documents for the top-10 results were selected. The queries are processed to create two permutations of the ordering of the documents.  The first permutation is a ranked list where the documents are sorted by their Statistical Language Model ({\em SLM}) score, which we will label as the {\em Samtla} order queries.  The second permutation is generated by shuffling the position of the documents, which we refer to as the {\em random} order queries. Each user completed 10 queries in the {\em Samtla} order, and the remaining 40 queries in {\em random} order. \\
\\
We measure system performance using the {\em random} order queries exclusively. The documents are sorted by their {\em SLM} score to recreate the {\em SLM} ranking, which we compare to the user-generated ranking, which we call the {\em consensus} ranking. The {\em consensus} ranking is created by aggregating all users' relevance grades for each document where ``Not relevant''$=1,... , $ ``Very relevant''$=4$.  We test for a presentation bias \cite{Barilan2009} by comparing this {\em consensus} ranking to the {\em display} order of the documents using both the full set of 50 queries and the 40 {\em random} order queries.  If users are influenced by the presentation order of the documents, we will find documents at the top of the {\em display} ranking being assigned higher relevance grades simply due to their position in the ranked list, which may actually be in lower positions according to the {\em SLM} ranking. A display bias will be apparent if there is a notable difference in the average scores across the two query sets. When discussing the performance measures we let $r_1$ and $r_2$ denote the \emph{system} ranking and {\em consensus} ranking respectively.
\subsubsection{User Interface}
The evaluation interface represents a cut-down version of the Samtla system, which isolates the search result window. At the top of the page we display the query submitted to the system, which was used to generate the list of top 10 search results.  Each document in the ranked list is displayed with the title and a short snippet showing the top three fragments containing the highlighted query in the document. Alongside each document is a drop-down box where the user selects an appropriate relevance grade.
\begin{figure}[!tbh]
	\centering
		\fbox{\includegraphics[width=\columnwidth]{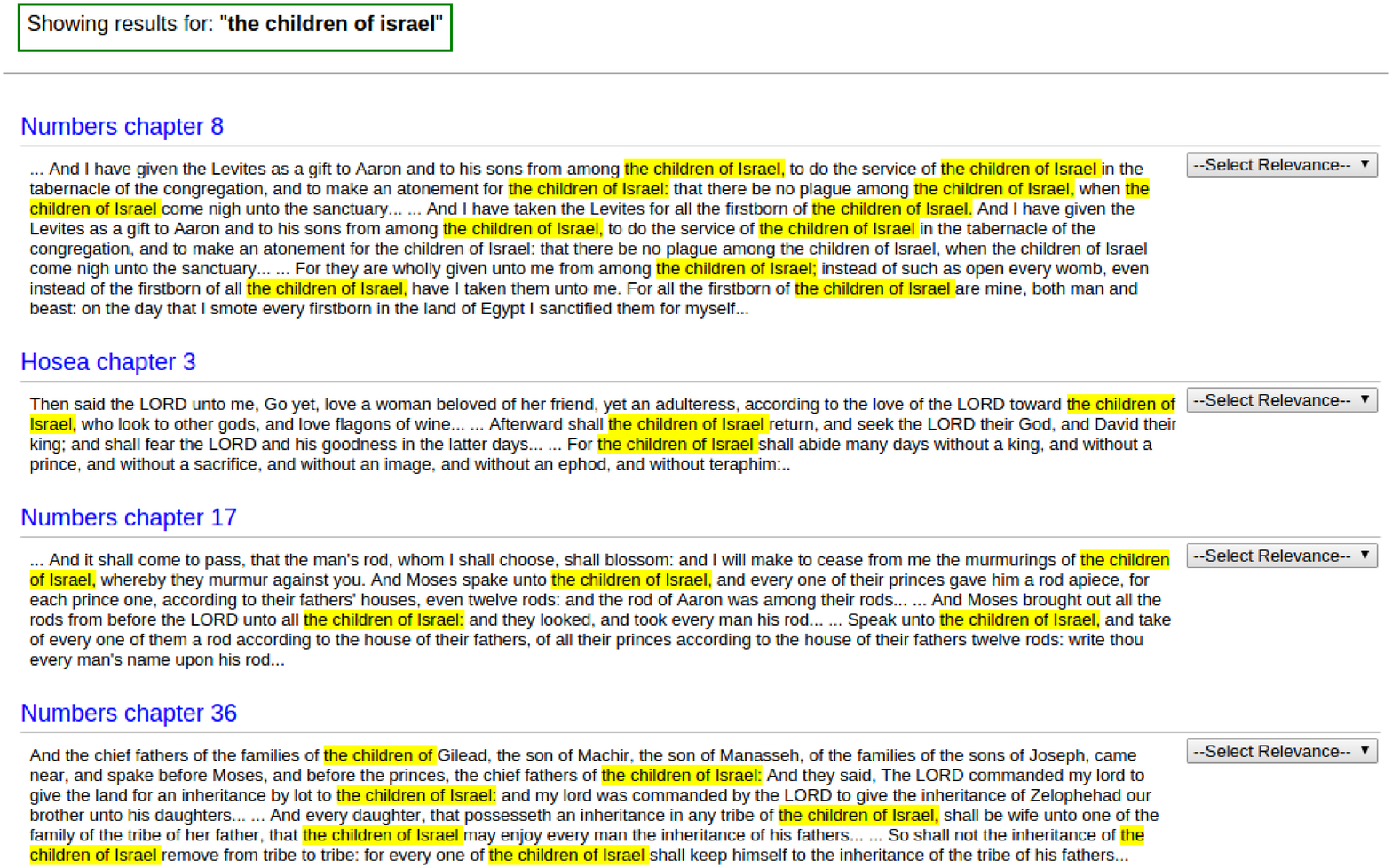}}
    	\caption{The evaluation page showing a single test.}
	    \label{fig:evalAMT}		
\end{figure}
\subsubsection{Selecting participants}
{\em Prolific Academic} provides a number of filters to enable researchers to exclude certain users based on specific attributes stored as part of the users profile. The main filtering criteria applied to our study was to ensure that users were fluent English speakers.  When we speak of a participant it should be clear from the start that they represent a member of the public and are not necessarily concerned with the motivation behind the specific study, or that they have a background in the type of data you are presenting to them as a researcher.  Therefore it is important to prepare for this fact and attempt to filter the crowd of individuals for those who will be competent in completing the required task.\\
\\
As mentioned above, the evaluation contained two test queries at the beginning of the survey.  Not all users will have read or understood the instructions \cite{GarciaMolina2016}, and others may simply assign relevance grades at random in order to complete the survey and receive payment as quickly as possible, known as "gaming" the system \cite{Kittur2008}. It is important to plan and mitigate against these types of user behaviour, especially when it comes to crowdsourcing, since it is generally not feasible to monitor the performance of users in realtime during the evaluation (although see \cite{Zaidan2011}). \\
\\
The first test query contained the top-5 ranked documents for the single word query ``Satan'' displayed at the top of the page.  The remaining 5 entries of the search results contained the snippets from a completely different, much longer query, ``chief priests and scribes''.  To pass the test, the user has to assign ``Not Relevant'' to these last five documents since they do not match the query ``Satan''.  \\
\\
The second test query ``Jesus Christ'' was composed of the top-10 documents ranked in reverse order of relevance.  In order to continue on to the evaluation, the user must assign higher relevance grades to documents as the rank position increases.  The results presented in the next section demonstrate that test queries are an important design consideration.
\subsection{Evaluation Measures}
We adopt two sets of measures for calculating the system performance. The first set of measures assesses the correlation between the system and the user generated ranking over each query in the {\em SLM} rank and {\em display} rank order.  If the system ranking is highly correlated with the user ranking then we can conclude that the ranking performance of the system closely matches that of a human assessor.  The second measure evaluates the ranking quality of the system using the Normalised Discounted Cumulative Gain measure (NDCG), which is commonly adopted for graded-relevance based evaluations \cite{Jarvelin2002}.  We perform the measures over both {\em SLM} and {\em display} permutations. In the following subsections, we describe each of the measures in more detail, before presenting a summary of the final results.
\subsubsection{Correlation Measures}
We measure the degree of correlation between the system and the users with {\em Spearman's footrule} \cite{diaconis1977} and the {\em M}-measure variant \cite{Levene2006}. These non-parametric measures describe the degree of correlation between two ranked lists, and provide similar results to other correlation measures including Spearman's $\rho$ and Kendall's $\tau$ \cite{Fagin2003}.  In our case the two ranked lists are represented by the {\em SLM} ranking $r_{1}$, and the user {\em consensus} ranking $r_{2}$.  We discuss each of the correlation measures in more detail, where we will abbreviate {\em Spearman's footule} to simply {\em Footrule} throughout the rest of this section.  \\
\\
The {\em Footrule} is calculated by summing the result of the absolute differences between the rank positions of the documents for each individual ranked list. The \emph{Footrule}, denoted by $Fr$, is more formally defined as follows:
\begin{equation}
Fr(r_{1}, r_{2}) = \sum_{i=1}^{k}|(r_{1}(i) - r_{2}(i))|
\label{eq:spearmanF}
\end{equation}\\
where $r_{1}$ and $r_{2}$ are two ranked lists assumed to contain the same set of documents, and $k$ is the size of the ranked list, in our case $k=10$, which represents the top-10 ranked documents. In order to use the {\em Footrule} as a metric, we need to normalise the result by calculating the maximum possible value, through:
\begin{equation}
F = 1 - \frac{Fr(r_{1}, r_{2})}{maxFr(k)}
\end{equation}\\
where $maxFr$ represents the maximum value, which when $k$ is an even number $maxFr = \frac{1}{2}k^2$, and if $k$ is an odd number then $maxFr = \frac{1}{2}(k + 1)(k-1)$. This ensures the resulting {\em Footrule} falls in the range of 0 and 1 where a value close to 1 means that the two ranked lists are highly similar.\\
\\
When evaluating search results, however, we may wish to consider the fact that documents in the top ranks are often considered the most relevant to the users information need than documents appearing in lower ranks \cite{Levene2006}.  To give more weight to the top ranked documents, we apply the {\em M}-measure, which was designed to place more emphasis on ranked lists containing identical or near-identical sets of documents in the top rank positions. Due to the fact that the ranked lists contain the same set of documents, we can drop the terms $S$ and $T$ mentioned in \cite{Levene2006}, which record the set of documents unique to $r_{1}$ and $r_{2}$, respectively, and reformulate the {\em M}-measure more precisely as:
\begin{equation}
\label{eq:m_measure}
m = \sum_{i=1}^{k} {|\frac{1}{r_{1}(i)} - \frac{1}{r_{2}(i)}|},
\end{equation}\\
where we calculate the sum of the absolute differences between each document's {\em SLM} rank and {\em consensus} rank. Next we calculate the maximum value $max$ {\it M}, which is defined through:
\begin{equation}
\label{eq:max_ofm}
max {\it M} = \sum_{i=1}^k{|\frac{1}{i} - \frac{1}{k-i+1}|}
\end{equation}
\\
Lastly, we normalise $m$ by deducting 1 from the result of the division of $m$ by the maximum value  $maxM$ to obtain a metric ranging between 0 and 1: \\
\begin{equation}
\label{eq:max_m_measure}
M = 1-\frac{m}{max {\it M}(k)}
\end{equation}
\\
The average by query is calculated by summing the scores for each correlation measure, {\em Footrule} and {\em M}-measure, and then dividing the result by the total number of queries.  We repeat this process for each user by first gathering the per query correlation scores for each user and then dividing the result by the total number of queries, and then take a further average for each user. The average {\em user agreement} represents the degree of correlation between each user and the {\em consensus} ranking, in other words, the {\em user agreement} describes the average correlation between an individual user and what we could consider to be the ``wisdom'' or ``opinion'' of the crowd.  We produce a {\em consensus} ranking for each query, and measure the correlation between this ranking and the individual user ranking for the given query.  Next, we calculate the average correlation per query for each user, as we did before, and then compute the average consensus based on the per user averages.
\subsubsection{Normalised Discounted Cumulative Gain (NDCG)}
While the correlation measures tell us how well the system generated ranking correlates with the user judgements, it does not directly describe the quality of the ranking algorithm. The ranking performance of an information retrieval system can be measured with the {\em Normalised Discounted Cumulative Gain} ({\em NDCG}) computed over the graded relevance scores.  The {\em $NDCG$} for a ranked list of size $k$ is calculated through:
\begin{equation}
NDCG_{k} = \frac{DCG_{k}}{IDCG_{k}}
\end{equation}\\
where $DCG$ is the discounted cumulative gain, $IDCG$ represents the ideal $DCG$, obtained by sorting the documents in descending order by relevance value, and then calculating the $DCG$ to get the maximum $DCG$, which is used in the normalisation step. \\
\\
We selected two discounting functions for comparison. The first method reduces the contribution of the relevance score according to rank position, which we define as $n$. The second discounting function is the more common approach where the relevance score is discounted by $\log_2$ of the rank position.  The $DCG$ at a particular rank position $k$ is defined as:  \\
\begin{equation}
DCG_{k}(r) = rel_{1} + \sum_{i=2}^{k} \frac{rel_{i}}{\log_2 i}
\end{equation}
\linebreak
where $r$ is a ranked list containing documents, and $rel_{i}$ is the relevance score at position $i$ and $\log_2 i$ represents the discounting function.  The discounting function models user persistence \cite{Jarvelin2002} in terms of whether the user will continue to look for more documents further down the search results.  This is achieved by reducing the contribution of the relevance score assigned to each document as a function of its position in the ranked list.  Documents appearing later in the ranked list are unlikely to be as relevant to the user as those in the top ranks and therefore only a small part of the document relevance score is passed on to the cummulative gain. The resulting $NDCG$ score ranges from 0 to 1, where a value of 1 means the ranking quality of the system is perfect as it is equivalent to the $IDCG$. We take the query results for each user and calculate the $NDCG$ and an average for each user, before calculating a final average over all queries.  
\subsubsection{Significance testing}
\label{sec:bootstrap}
As part of the assessment we evaluate the statistical significance of the results.  We adopt the {\em bootstrap} method \cite{davisonHinkley1997, Smucker2007, Sakai2006}, which attempts to approximate the original underlying distribution of the population, by selecting a series of random samples of size $N$ with replacement from the observed population data. An advantage of the bootstrap method is that it is compatible with any statistical measure \cite{Smucker2007}, meaning we can use the correlation and {\em NDCG} scores as our test statistics.  Under the bootstrap method, we assume that the null hypothesis is that there is no difference between the ranking generated by the system and the ranking generated by the user evaluations.  The difference is considered significant, with respect to the stated significance level, if the confidence intervals do not overlap.  In order to obtain the confidence intervals we generate a series of samples by selecting a value at random from the original measures ({\em Footrule}, {\em M}-measure, and $NDCG$) to generate a sample equivalent in size to the number of queries or users in the original evaluation.  The sampling process can be thought of as extracting values from the rows and columns of a $n$ by $m$ matrix, where the rows contain the correlation or $NDCG$ scores by query, and the columns represent the per user scores.  Each random sample $b$, where $b=1, ..., B$, is composed of values selected with replacement. We perform this operation for a total sample size of $B=1000$ and calculate the average of the test statistic for each sample. Calculating the final confidence interval then involves sorting the averages in ascending order, and selecting the values that fall at the $B(1-(\alpha/2))$ and $B(\alpha/2)$ percentile, where $\alpha$ is the required significance level and $\alpha=0.05$ represents a 95\% confidence interval.  We take an average over the lower and upper bounds of the confidence intervals and partition the results by query, user, and {\em user agreement} (see Section~\ref{sec:finalresults}).
\subsection{Evaluation Results}
\label{sec:finalresults}
The evaluation was run over several days and 65 participants attempted the evaluation.  A total of 24 users successfully completed the survey and the majority of the submissions were received from men between the ages of 20-30 years, and resident or born in North America. Of the toal submissions received, we excluded 10 users due to incomplete results caused by connection timeout issues, and 31 users who failed to pass the test queries, which is almost half of the total attempted submissions. It is interesting to note that 10 of the users who failed the test did not even pass the first test query.  This means they were unable to identify that the top-10 results were composed of two completely different queries of different lengths (a short query versus a long verbose query), which highlights the importance of designing tests as part of an evaluation to filter potentially poor performing users.  
\subsubsection{Correlation measures}
In this section we report the final correlation scores. Before continuing, we establish a baseline figure for each measure.  We compute the {\em Footrule} and {\em M}-measure between the {\em SLM} ranking and the {\em display} order of the documents for each query, and then take the average over all queries to obtain an average baseline score. The baseline figures are reported in Table~\ref{table:corr_base}.\\
\begin{table}[!htbp]
\centering
\begin{tabular}{|r|c|c|c|c|c|c|c|c|c|c|c|c|c|c|c|c|}
\hline
\multicolumn{3}{|c|}{ Baseline {\it Correlation}} \\
\hline
{\it Type} & {\it Footrule} & {\it M-measure}\\ \hline
{\it Random} queries (40) & 0.400 & 0.369 \\ \hline
\end{tabular}
\caption{Baseline values for each correlation measure divided by query type}
\label{table:corr_base}
\end{table}
The baseline correlation for the 10 Samtla queries is 1.0 since the display and the system ranking are equivalent and is not included here. In terms of the baseline for the 40 {\em random} order queries, we can see that it is quite low across the two measures, but does show that there is some correlation despite the shuffling process. The final results for each measure are displayed in Table~\ref{table:corr_results1} and Table~\ref{table:corr_results2} where we present the average {\em Footrule} and {\em M}-measure for the {\em SLM} and {\em display} ranking compared to the user {\em consensus} ranking, respectively. For each form of analysis, we divide the results into query, user, and user consensus averages, and report the 95\% confidence interval in square brackets, obtained from the bootstrap (see Section~\ref{sec:bootstrap}).\\
\begin{table}[!htbp]
\centering
\begin{tabular}{|r|c|c|c|c|}
\hline
& \multicolumn{2}{|c|}{{\bf SLM}} \\
\hline
{\it Samtla} queries (10) & {\it Footrule} & {\it M-measure}\\ \hline
Query & 0.775 [0.775 - 0.779] & 0.840 [0.840 - 0.844]\\ \hline
User & 0.863 [0.859 - 0.863]  & 0.906 [0.903 - 0.906]\\ \hline
User consensus & 0.800 [0.795 - 0.798] & 0.847 [0.845 - 0.848]\\ \hline
\hline
{\it Random} queries (40) & {\it Footrule} & {\it M-measure}\\ \hline
Query & 0.757 [0.756 - 0.759]& 0.761 [0.759 - 0.763]\\ \hline
User & 0.853 [0.851 - 0.854]& 0.846 [0.844 - 0.847]\\ \hline
User consensus & 0.716 [0.715 - 0.718] & 0.737 [0.735 - 0.739]\\ \hline
\end{tabular}
\caption{Average correlation scores for the {\em random} queries ordered by the {\em SLM} ranking divided into query, user, and user consensus}
\label{table:corr_results1}
\end{table}
\\
The results of Table~\ref{table:corr_results1} show that the user relevance judgments for the 10 Samtla queries is comparable or higher for the Footrule and the M-measure, implying that a display bias may be present. In particular, we see that the average score by user for the {\em M}-measure is 0.906, suggesting that users were more likely to assign higher relevance grades to a selection of documents appearing at the top of the search results.  Consequently, we discard this set of queries from the analysis, and focus solely on the {\em random} order queries for the remainder of this section. \\
\\
The average correlation scores for the 40 {\em random} queries are positively correlated to the {\em consensus} ranking.  Looking at the {\em Footrule}, we see that the query average of 0.757 is lower than the user average of 0.853. This may be attributed to query length, in terms of the fact that longer queries had an average of 3 to 4 highly relevant documents, with the remaining documents containing only partial matches. On the other hand, short keyword queries tended to have full matches to the query in all the documents retrieved and so users assigned proportionally higher relevance grades across more documents in the search results. We also observe similar results for the average {\em M}-measure, which shows that users were assigning more relevance to a selection of documents that are ranked in the top positions according the underlying $SLM$. \\
\\
The average {\em user consensus} also suggests that the majority of users were in agreement with the crowd opinion of which documents were the most relevant.  Turning to the correlation scores calculated over the {\em display} ranking, the results in Table~\ref{table:corr_results2} summarise the final averages for the {\em Footrule} and {\em M}-measure over the 40 {\em random} order queries according to the order in which the documents were presented to the users.\\
\\
\begin{table}[!htbp]
\centering
\begin{tabular}{|r|c|c|c|c|c|c|c|}
\hline
{\it Random} queries (40) & {\it Footrule} & {\it M-measure}\\ \hline
Query &  0.402 [0.400 - 0.404]& 0.474 [0.471 - 0.476] \\ \hline
User &  0.435 [0.434 - 0.437]& 0.416 [0.415 - 0.417]\\ \hline
User consensus & 0.660 [0.658 - 0.661] & 0.669 [0.668 - 0.671]\\ \hline
\end{tabular}
\caption{Average correlation scores for the {\em display} ranking divided into query, user, and average user consensus for the {\em random} order queries.}
\label{table:corr_results2}
\end{table}
We can see that the correlation is much lower than both the 10 {\em Samtla} queries and the {\em random} order queries ranked according to the $SLM$ (Table~\ref{table:corr_results1}) across queries and users.  This suggests that the users were less sensitive to the presentation order of the documents, in other words they were attempting to do a good job rather than assigning relevance as a function of the document position.  The average correlation scores for the {\em display} order of the 40 {\em random} order queries are still positively correlated, suggesting that there is some presentation bias. However, if we take in to account the degree of correlation that already existed between the two permutations on the order of documents, that is 0.400 and 0.369 for the {\em Footrule} and {\em M}-measure, respectively (see Table~\ref{table:corr_base}), then we could argue that the average correlation scores are actually much smaller.  Taking this issue in to account, we can conclude that there was not much bias in the users' judgements in terms of the presentation order of the documents.\\
\\
To summarise, there is an observable difference in the correlation scores for the {\em SLM} order and the {\em display} order, with the correlation measures for the {\em SLM} being higher than the {\em display} order suggesting that users agreed with the ranking generated by the {\em SLM}.  In general, it appears that the crowd of users were not affected by the presentation order of the documents when they were in {\em random} order, and we are able to reject the null hypothesis that there is no correlation between the {\em SLM} system ranking and the ranking generated from the user relevance judgements, which is supported by the fact that the confidence intervals do not overlap meaning we have a significant result at the 95\% confidence level.  
\subsubsection{Normalised Discounted Cumulative Gain (NDCG)}
Before presenting the results for the $NDCG$ measure, we establish a baseline once again establish a baseline. This is done using a different method to the correlation measures, where we simulate the input provided by 1000 random users.  Each user is represented by a random assignment of relevance grades to the documents for each query, which we then summarise by computing the average $NDCG$ by user and query for each discounting function $n$ and $log_{2}$, respectively, which are presented in Table~\ref{table:baseline_NDCG} below.\\
\begin{table}[!htbp]
\centering
\begin{tabular}{|r|c|c|c|c|c|c|c|c|c|c|c|c|c|c|c|c|}
\hline
\multicolumn{2}{|c|}{ Baseline {\it NDCG}} \\
\hline
{\it n} & {\it $log_2$}\\ \hline
0.853 & 0.870\\ \hline
\end{tabular}
\caption{Baseline NDCG}
\label{table:baseline_NDCG}
\end{table}
\\
The average baseline figures are fairly close to the maximum $NDCG$, with the logarithmic discounting function $log_2$ being slightly less aggressive than the discounting by rank position $n$. We have found that we obtain similar results regardless of the adopted discounting function, however, we include both as they provide different models of user persistance, with $n$ representing a more impatient user. The final average scores for the $NDCG$ applied to the {\em SLM} ranking and the {\em display} ranking are presented below (see Table~\ref{table:ndcg@10_a} and Table~\ref{table:ndcg@10_b}). We make a distinction between the {\em display} ranking for the 10 {\em Samtla} queries and the 40 {\em random} order queries and report the total average by query and user with their 95\% confidence intervals presented alongside in square brackets. \\
\begin{table}[!htbp]
\centering
\begin{tabular}{|r|c|c|c|c|c|c|c|c|c|c|c|c|c|c|c|c|}
\hline
NDCG@10 & \multicolumn{2}{|c|}{{\bf SLM}}\\
\hline
{\it Samtla} queries (10) & {\it n} & {\it $log_2$}\\ \hline
Query & 0.985 [0.985 - 0.985] & 0.988 [0.987 - 0.988] \\ \hline
User & 0.985 [0.985 - 0.985] & 0.987 [0.987 - 0.987]\\ \hline
\hline
{\it Random} queries (40) & {\it n} & {\it $log_2$}\\ \hline
Query & 0.981 [0.980 - 0.981]& 0.983 [0.983 - 0.984]\\ \hline
User & 0.982 [0.981 - 0.982]& 0.984 [0.983 - 0.984]\\ \hline
\end{tabular}
\caption{Average correlation scores for the {\em SLM} ranking divided in to query and user averages with the 95\% confidence intervals in square brackets.}
\label{table:ndcg@10_a}
\end{table}\\
As with the correlation scores, we can see that the users tended to assign higher relevance to the top documents in the search results, as illustrated by the $NDCG$ scores being very close to the $IDCG$.  The query and user averages based on discounting by $n$ are equal as a result of rounding, but there was only a slight difference between the two (0.98573 and 0.98564 respectively).  As mentioned, we can see that the $NDCG$ is slightly higher for the 10 Samtla queries than the 40 random queries, suggesting that users assigned higher scores to the top documents as a result of their position being at the top of the search results. Consequently, we remove these 10 queries from the results and discussion, as there would once again appear to be a slight presentation bias.\\
\\
Turning to the average $NDCG$ scores for the {\em SLM} ranking of the 40 {\em random} order queries, we see that the average query and user $NDCG$ scores are quite close for the {\em SLM} ranking (0.983 by query, and 0.984 by user assuming a discounting function of $log_{2}$). If we compare these results to the $NDCG$ scores of {\em display } order of the queries (see Table~\ref{table:ndcg@10_b}) and the baseline $NDCG$ scores (see Table~\ref{table:baseline_NDCG}), there was less of a presentation bias due to the relatively low $NDCG$ at 0.881 for the discounting function by rank position $n$, and 0.894 for the discounting function of $log_{2}$.  This means that users were not heavily influenced by the presentation order of the documents. We observe that the 95\% confidence intervals from the {\em bootstrap} process do not overlap, which means we can conclude that the results are significant at the $\alpha=0.05$ level.  And we noted that the users gave more relevance to documents appearing in the top ranks of the {\em random} order queries, represented by the high average $NDCG$ score for the {\em SLM} ranking (Table~\ref{table:ndcg@10_b}). Naturally, these results assume that we take in to account the baseline $NDCG$ scores, which means the $NDCG$ for the {\em display} order is actually much lower.\\
\\
\begin{table}[!htbp]
\centering
\begin{tabular}{|r|c|c|c|c|c|c|c|c|c|c|c|c|c|c|c|c|}
\hline
NDCG@10 & \multicolumn{2}{|c|}{{\bf Display}} \\
\hline
{\it Random} queries (40) & {\it n} & {\it $log_2$}\\ \hline
Query & 0.881 [0.881 - 0.883] & 0.894 [0.894 - 0.896] \\ \hline
User & 0.882 [0.881 - 0.883] & 0.895 [0.894 - 0.896]\\ \hline
\end{tabular}
\caption{Average correlation scores for the{\em display} ranking divided in to query and user averages for the 40 {\em random} order queries, with the 95\% confidence intervals in square brackets.}
\label{table:ndcg@10_b}
\end{table}
To summarise, on the basis of the performance measures presented above, we can say that users were highly correlated with the {\em SLM} order of the queries than the {\em display} order when analysing the results of the {\em random} order queries independently of the 10 {\em Samtla} queries.  The users were more influenced by the presentation order of the 10 {\em Samtla} queries, in the sense that they were slightly more generous with their relevance grades, where they tended to assign higher relevance to a few documents at the very top of the search results shown by the high $M$-measure and $NDCG$ scores. Out of the 50 queries completed by each user, 80\% of them were presented in random order, yet we see that the users consistently assigned more relevance to the documents that received the highest document score according to the underlying $SLM$, and we can see that these scores are not the result of users assigning relevance at random, or "gaming" the system, in part due to the role played by the quality assessment represented by the test queries.  We also observed users revisiting their earlier relevance assignments, when they encountered highly relevant documents at the bottom of the result page, caused by the random shuffle process.  Therefore, there is significant evidence to suggest that users were attempting to do a good job and were not assigning relevance grades purely at random, but based on what they considered to be relevant given the provided query context.  \\
\\
We can conclude then, that the ranking quality of {\em Samtla} and its underlying $SLM$ correlates well with the ranking generated by the user relevance judgements, both in terms of which documents were relevant and also of the top document, which were most likely to meet their information across query types, from single word queries to long  more verbose queries.
\subsection{Discussion}
\begin{figure}[!tbh]
	\centering
		\fbox{\includegraphics[width=\columnwidth]{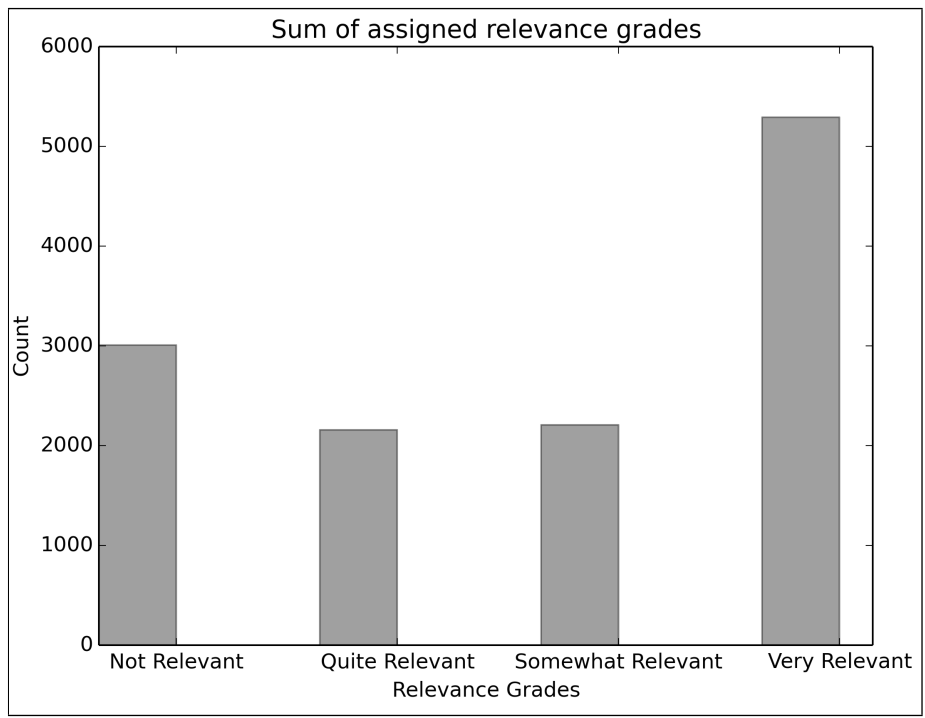}}
    	\caption{Distribution of relevance grades used in the evaluation.}
	    \label{fig:evalrelevance}		
\end{figure}

Crowdsourcing has its challenges, in particular, the researcher has little control over the evaluation process once it is launched and available online.  Therefore, as we have demonstrated, it is necessary to consider the use of test queries in order to filter out bad users upfront e.g. those who have not understood the task or do not have the correct attitude.  This increases the quality of the submissions, and mitigates against issues that can arise, such as an unhappy user as a result of a rejected submission, or withholding payment due to a suspect submission. These issues can be difficult to resolve and may have an impact on your reputation, and consequently on whether you will be able to submit future evaluations with the same crowd sourcing platform.\\
\\
The design of the evaluation should record data that permits the testing of a display bias, since some users may assign relevance to document in the top ranks without necessarily digesting the snippets fully.  This is easily achievable by randomising the order of the queries.  It is also worth recording a timestamp for each response. This enables the researcher to check for users who are speeding through the evaluation at a rate that exceeds the ability to comfortably digest the information related to the task.  We found that users assigned relevance at an average rate of three seconds per rank position.  The minimum time taken was one second, which we could argue is not enough time to digest the snippet and then navigate to the drop-down box to select a relevance grade.  The maximum time to select a relevance grade was 13 minutes, but this is likely the result of users being interrupted or distracted from the task.  \\
\\
Furthermore, the difference between the total query average and user averages can be explained by the fact that users tended to adopt their own strategy for assigning relevance.  A large number of users did not make use of all relevance grades (see Figure~\ref{fig:evalrelevance}), but instead adopted a binary relevance approach where they only assigned grades of ``Very Relevant'' or ``Not Relevant'' to the documents.  The short queries tended to have more relevant documents in the top-10 meaning that the user tended to judge relevance based on the total number of highlighted terms in the snippet.  On the other hand, the longer verbose queries contained an average of 3 to 4 ``Very Relevant'' documents, with the remaining results containing partial matches to the query, which received less relevance. For example, documents containing a full match for the query ``...$[$As the Lord commanded$]$...'' naturally received higher relevance scores than the partial match ``...$[$As th$]$y $[$Lord commanded$]$...''. However, this is often user-dependent, and it could be argued that a researcher of the Bible would find the latter example just as relevant to their information need, or at least, that it provides an interesting example for their research.\\
\\
In conclusion, we have shown that non-parametric correlation and $NDCG$ measures provide a good basis for assessing the performance of an information retrieval system.  The non-parametric correlation measures show the degree of agreement between what users considered relevant and the ranking generated by the $SLM$ (see Section~\ref{sec:datamodel}). On the other hand, the $NDCG$ described the ranking quality of the ranked lists, and we observe that the system consistently produces a ranking where the top ranks are occupied by the most relevant documents. We also described how we can measure the overall opinion or agreement between the users by comparing each user with the {\em consensus} ranking, which showed that each individual user agreed on average with the ranking generated by the crowd. Lastly, the significance of the results was evaluated with the {\em bootstrap} method, which is non-parametric, relatively simple to implement, and as effective as other significance tests \cite{Sakai2006}.  \\
\\
Using crowdsourcing as a platform for system evaluation provides researchers with access to a large group of potential participants, but as we have demonstrated, it is necessary to design the evaluation in such a way so as to minimise technical challenges, minimise poor quality results, and record data on user interaction with the evaluation software in order to spot potential cheating.
\section{Concluding Remarks}
\label{sec:conclusion}
We have introduced the underlying framework of the Samtla system (Section~\ref{sec:architecture}), and the data structures and algorithms adopted. We showed how statistical language models can be used for ranking documents according to user queries (Section~\ref{sec:datamodel}) and demonstrated that our implementation is providing users with the most relevant documents in the top ranks of the search results (Section~\ref{sec:evaluation}).\\
\\
We also described how users interact with the system (Section~\ref{sec:UI}), and the tools we have currently released to our user groups (Section~\ref{sec:textmining}). The case studies provide an insight into how our users are currently using these tools to carry out their research (Section~\ref{sec:casestudy}).  \\
\\
We are now focusing on the development of the underlying framework where we look at additional parameters that can be incorporated in to the data model (see Section~\ref{sec:datamodel}) in order to add a layer of semantics to the search component. For example, we currently assume a uniform prior for all document probabilities when ranking the documents in response to a query. We can use the {\em JSD} matrix generated for the related documents tool (see Section~\ref{sec:Relateddocuments}) to compute a non-uniform prior, which will enable us to integrate document-specific knowledge as part of the Samtla query model.  Further work is centered on simple methods for identifying important events in the collection documents, which could be presented to users as a timeline.  This task is often referred to as event tracking and identification \cite{Sayyadi2009}. \\
\\
The main novelty of Samtla is the underlying probabilistic model that has enabled us to develop a diverse range of tools that are language independent and applicable to many document collections, including flexible search and mining of text patterns, document comparison, query and document recommendation, and the way the system can incorporate external sources of information in the form of metadata provided by users or third-party sources such as Wikipedia, to supplement the toolset.  Samtla aims to complement existing methods in the digital humanities by helping researchers with their research needs by providing a general purpose environment. \\
\\
In summary, we have discussed how systems developed for the Humanities can be made 'future-proof' in the sense of providing a generalised framework that can be easily extended to new document collections without changes to the underlying system components in order to compensate for language-specific issues such as word stemming and tokenisation.  Although Samtla is still in development we already have a number of Samtla systems available for a range of document collections (King James Bible, Aramaic Magic Bowls, Vasari, the Microsoft Corpus, and the Financial Times), which cover a broad range of corpora composed of one or more languages including Aramaic, Syriac, Mandaic, Hebrew, English, German, French, Hungarian, Italian, and Russian.  In addition, Samtla is not necessarily restricted to historic document collections, but can be extended straightforwardly to other application domains, which require search and mining of text patterns, such as medical and legal text collections.
\bibliographystyle{abbrv}
%\setcitestyle{authoryear,open={((},close={))}}

\end{document}